\documentclass{llncs}

\usepackage[utf8]{inputenc}
\usepackage[english]{babel}

\usepackage{amsmath,amsfonts,amssymb}
\usepackage{array}
\usepackage{url}

\usepackage{versions}
\excludeversion{FORABSTRACT}
\includeversion{FORREPORT}
\excludeversion{FORJOURNAL}
\def\pFORREPORT#1{\processifversion{FORREPORT}{#1}}
\def\pFORJOURNAL#1{\processifversion{FORJOURNAL}{#1}}

\newif\ifwithplots
\withplotstrue 

\usepackage{tikz,pgfplots}

\tikzset{
  baseline=(current bounding box.center)
}

\pgfplotscreateplotcyclelist{mycolor}{%
  red, every mark/.append style={solid,scale=0.8}, mark=o \\%
  blue, every mark/.append style={solid,scale=0.7}, mark=square \\%
  green!70!black, every mark/.append style={solid,scale=0.95}, mark=diamond \\%
  orange!40!red, every mark/.append style={solid,scale=0.95}, mark=x, dashed \\%
  teal, every mark/.append style={solid,scale=0.95}, mark=asterisk, dashed \\%
  brown!70!black, every mark/.append style={solid,scale=0.95}, mark=triangle, dashed \\%
  gray, every mark/.append style={solid,scale=0.8}, mark=o, densely dotted \\%
  purple!80!blue, every mark/.append style={solid,scale=0.7}, mark=square, densely dotted \\%
}

\pgfplotscreateplotcyclelist{myblack}{%
  black, every mark/.append style={solid,scale=0.8}, mark=o \\%
  black, every mark/.append style={solid,scale=0.7}, mark=square \\%
  black, every mark/.append style={solid,scale=0.95}, mark=diamond \\%
  black, every mark/.append style={solid,scale=0.95}, mark=x, densely dashed \\%
  black, every mark/.append style={solid,scale=0.95}, mark=asterisk, densely dashed \\%
  black, every mark/.append style={solid,scale=0.95}, mark=triangle, densely dashed \\%
  black, every mark/.append style={solid,scale=0.8}, mark=o, densely dotted \\%
  black, every mark/.append style={solid,scale=0.7}, mark=square, densely dotted \\%
}

\pgfplotsset{
  major grid style={thin,dotted,color=black!50},
  minor grid style={thin,dotted,color=black!50},
  grid,
  cycle list name={mycolor},
  every axis/.append style={
    line width=0.5pt,
    tick style={
      line cap=round,
      thin,
      major tick length=4pt,
      minor tick length=2pt,
    },
  },
  legend cell align=left,
  legend style={
    /tikz/every even column/.append style={column sep=3mm,black},
    /tikz/every odd column/.append style={black},
  },
  plotSpeedupMini/.style={
    width=67mm,height=49mm,
    xlabel near ticks,
    ylabel absolute=true,
    every axis y label/.append style={yshift=-13pt},
    max space between ticks=18pt,
    title style={yshift=-3pt},
  },
  plotSpeedup64/.style={
    plotSpeedupMini,xtick={1,8,16,32,48,64},
  },
  plotSpeedup48/.style={
    plotSpeedupMini,xtick={1,6,12,24,36,48},
  },
  plotSpeedup16/.style={
    plotSpeedupMini,xtick={1,2,4,6,8,12,16},
  },
  plotSpeedup8/.style={
    plotSpeedupMini,xtick={1,...,8},ytick={0,...,5},
  },
}

\def\SpeedupLegend{%
  \legend{pS$^5$-Unroll,pS$^5$-Equal,pMultikeyQuicksort,pRadixsort 8-bit,pRadixsort 16-bit,pRadixsort Akiba,pMKQS-SIMD Rantala,pMergesort-2way Rantala}%
}

\usepackage{etoolbox}
\makeatletter
\let\llncs@addcontentsline\addcontentsline
\patchcmd{\maketitle}{\addcontentsline}{\llncs@addcontentsline}{}{}
\patchcmd{\maketitle}{\addcontentsline}{\llncs@addcontentsline}{}{}
\patchcmd{\maketitle}{\addcontentsline}{\llncs@addcontentsline}{}{}
\makeatother

\usepackage{hyperref}
\hypersetup{
  breaklinks=true,
  colorlinks=true,
  citecolor=blue,
  linkcolor=blue,
  urlcolor=blue,
  bookmarksnumbered,
  bookmarksopen,
  pdftitle={Parallel String Sample Sort},
  pdfauthor={Timo Bingmann, Peter Sanders},
  pdfsubject={parallel string sorting algorithms, super scalar string sample sort},
  pdfkeywords={parallel string sorting, parallel algorithm, string sorter, sample sort},
}

\hypersetup{pdfpagescrop={92 62 523 748}}

\usepackage{breakurl}

\usepackage{amsfonts}

\newcommand{\ceil}[1]{\left\lceil #1\right\rceil}

\newcommand{\set}[1]{\left\{ #1\right\}}
\newcommand{\gilt}{:}

\newcommand{\realrange}[2]{\left[#1, #2\right]}

\newcommand{\unitrange}[2]{\realrange{0}{1}}

\newcommand{\Oh}[1]{\mathcal{O}\!\left( #1\right)}
\newcommand{\oh}[1]{\mathrm{o}\!\left( #1\right)}
\newcommand{\Th}[1]{\Theta\!\left( #1\right)}

\newcommand{\Om}[1]{\Omega\!\left( #1\right)}

\newcommand{\llabel}[1]{\label{\labelprefix:#1}}
\newcommand{\labelprefix}{} 

\newcommand{\discussionsize}{\small}

\newcommand{\frage}[1]{{\sf[ #1]}\marginpar{?}}

\newcommand{\punkt}{\enspace .}

\newenvironment{code}{\noindent
\begin{tabbing}%
\hspace{2em}\=\hspace{2em}\=\hspace{2em}\=\hspace{2em}\=\hspace{2em}\=%
\hspace{2em}\=\hspace{2em}\=\hspace{2em}\=\hspace{2em}\=\hspace{2em}\=%
\kill}{\end{tabbing}}

\newcommand{\labelcommand}{}
\newcommand{\captiontext}{}
\newsavebox{\codeparam}
\newcounter{lineNumber}
\newenvironment{disscodepos}[3]{%
\renewcommand{\labelcommand}{#2}%
\renewcommand{\captiontext}{#3}%
\sbox{\codeparam}{\parbox{\textwidth}{#3}}%
\begin{figure}[#1]\begin{center}\begin{code}\setcounter{lineNumber}{1}}{%
\end{code}\end{center}\caption{\llabel{\labelcommand}\captiontext}\end{figure}}

{\end{disscodepos}}

\newdimen\endofsize\endofsize=0.5em
\def\endofbeweis{~\quad\hglue\hsize minus\hsize
                 \hbox{\vrule height \endofsize width
\endofsize}\par}

\newcommand{\lcp}{\mathrm{lcp}}
\newcommand{\Strings}{\mathcal{S}}

\renewcommand{\frage}[1]{}

\title{Parallel String Sample Sort}
\author{Timo Bingmann, Peter Sanders}
\institute{
Karlsruhe Institute of Technology, Karlsruhe, Germany\\
\email{\{bingmann,sanders\}@kit.edu}}

\begin{document}

\maketitle

\begin{abstract}
  We discuss how string sorting algorithms can be parallelized on modern
  multi-core shared memory machines. As a synthesis of the best sequential
  string sorting algorithms and successful parallel sorting algorithms for
  atomic objects, we propose string sample sort. The algorithm makes effective
  use of the memory hierarchy, uses additional word level parallelism, and
  largely avoids branch mispredictions. Additionally, we parallelize variants of
  multikey quicksort and radix sort that are also useful in certain situations.
\end{abstract}

\pagestyle{plain}

\section{Introduction}

Sorting is perhaps the most studied algorithmic problem in computer science.
While the most simple model for sorting assumes \emph{atomic} keys, an important
class of keys are strings to be sorted lexicographically. Here, it is important
to exploit the structure of the keys to avoid costly repeated operations on the
entire string.  String sorting is for example needed in database index
construction, some suffix sorting algorithms, or MapReduce tools. Although there
is a correspondingly large volume of work on sequential string sorting, there is
very little work on parallel string sorting. This is surprising since
parallelism is now the only way to get performance out of Moore's law so that
any performance critical algorithm needs to be parallelized. We therefore
started to look for practical parallel string sorting algorithms for modern
multi-core shared memory machines. Our focus is on large inputs. This means that
besides parallelization we have to take the memory hierarchy and the high cost
of branch mispredictions into account.

In Section~\ref{sec:prelim} we give an overview of string sorting
algorithms, acceleration techniques and parallel atomic sample sort. We then
propose our new string sorting algorithm super scalar string sample sort (S$^5$)
in Section~\ref{sec:s5}. The rest of Section~\ref{sec:smp-parallelize} describes
two competitors and we compare them experimentally in
Section~\ref{sec:experiments}. For all instances except random strings, S$^5$
achieves higher speedups on modern multi-core machines than our own parallel
multikey quicksort and radixsort implementations, which are already better than
any previous ones.

\begin{FORABSTRACT}
After introducing notation and previous approaches in Section~\ref{sec:prelim},
Section~\ref{sec:smp-parallelize} explains our parallel string sorting
algorithms, in particular super scalar string sample sort (S$^5$) but also
multikey quicksort and radix sort. These algorithms are evaluated experimentally
in Section~\ref{sec:experiments}.
\end{FORABSTRACT}

We would like to thank our students Florian Drews, Michael Hamann, Christian
Käser, and Sascha Denis Knöpfle who implemented prototypes of our ideas.

\section{Preliminaries}\label{sec:prelim}

Our input is a set $\Strings=\set{s_1,\ldots,s_n}$ of $n$ strings with total
length $N$.  A string is a zero-based array of $|s|$ characters from the
alphabet $\Sigma=\set{1,\ldots,\sigma}$. For the implementation, we require that
strings are zero-terminated, i.e., $s[|s|-1]=0 \notin \Sigma$.  Let $D$ denote
the \emph{distinguishing prefix size} of $\Strings$, i.e., the total number of
characters that need to be inspected in order to establish the lexicographic
ordering of $\Strings$. $D$ is a natural lower bound for the execution time of
sequential string sorting. If, moreover, sorting is based on character
comparisons, we get a lower bound of $\Omega(D+n\log n)$.

Sets of strings
are usually represented as arrays of pointers to the beginning of each
string. Note that this indirection means that, in general, every access to a
string incurs a cache fault even if we are scanning an array of strings. This is
a major difference to atomic sorting algorithms where scanning is very cache
efficient.
Let $\lcp(s,t)$ denote the length of the \emph{longest common
  prefix} (LCP) of $s$ and $t$. In a sequence or array of strings $x$ let $\lcp_x(i)$
denote $\lcp(x_{i-1},x_i)$.
Our target machine is a shared memory system supporting $p$ hardware threads
(PEs -- processing elements) on $\Th{p}$ cores.

\subsection{Basic Sequential String Sorting Algorithms}\label{sec:basic-sequential}

\emph{Multikey quicksort} \cite{BenSed97} is a simple but effective
adaptation of quicksort to strings.  When all strings in $\Strings$ have a
common prefix of length $\ell$, the algorithm uses character $c=s[\ell]$ of a
pivot string $s\in \Strings$ (e.g. a pseudo-median) as a \emph{splitter}
character. $\Strings$ is then partitioned into $\Strings_<$, $\Strings_=$, and
$\Strings_>$ depending on comparisons of the $\ell$-th character with
$c$. Recursion is done on all three subproblems. The key observation is that the
strings in $\Strings_=$ have common prefix length $\ell+1$ which means that
compared characters found to be equal with $c$ will never be considered
again. Insertion sort is used as a base case for constant size inputs. This
leads to a total execution time of $\Oh{D+n\log n}$. Multikey quicksort works
well in practice in particular for inputs which fit into the cache.

\emph{MSD radix sort} \cite{mcilroy1993engineering,ng2007cache,karkkainen2009engineering} with common prefix length $\ell$ looks at the $\ell$-th
character producing $\sigma$ subproblems which are then sorted recursively with
common prefix $\ell+1$. This is a good algorithm for large inputs
and small alphabets since it uses the maximum amount of information within a
single character. For input sizes $\oh{\sigma}$ MSD radix sort is no longer
efficient and one has to switch to a different algorithm for the base case. The
running time is $\Oh{D}$ plus the time for solving the base cases. Using
multikey quicksort for the base case yields an algorithm with running time
$\Oh{D+n\log\sigma}$. A problem with large alphabets is that one will get many
cache faults if the cache cannot support $\sigma$ concurrent output streams (see
\cite{MehSan03} for details).

\emph{Burstsort} dynamically builds a trie data structure for the input
strings. In order to reduce the involved work and to become cache efficient, the
trie is build lazily -- only when the number of strings referenced in a
particular subtree of the trie exceeds a threshold, this part is expanded. Once all
strings are inserted, the relatively small sets of strings stored at the leaves
of the trie are sorted recursively (for more details refer to \cite{sinha2004cache-conscious,sinha2007cache-efficient,sinha2010engineering} and the references therein).

\emph{LCP-Mergesort} is an adaptation of mergesort to strings that saves and
reuses the LCPs of consecutive strings in the sorted subproblems
\cite{ng2008merging}.

\subsection{Architecture Specific Enhancements}

\emph{Caching of characters} is very important for modern memory hierarchies as it reduces the number of cache misses due to random access on strings.
When performing character lookups, a caching algorithm copies successive characters of the string into a more convenient memory area.
Subsequent sorting steps can then avoid random access, until the cache needs to be refilled.
This technique has successfully been applied to radix sort \cite{ng2007cache}, multikey quicksort \cite{rantala07web}, and in its extreme to burstsort \cite{sinha2007cache-efficient}.

\emph{Super-Alphabets} can be used to accelerate string sorting algorithms which
originally look only at single characters.  Instead, multiple characters are
grouped as one and sorted together.  However, most algorithms are very sensitive
to large alphabets, thus the group size must be chosen carefully. This approach
results in 16-bit MSD radix sort and fast sorters for DNA strings. If the
grouping is done to fit many characters into a machine word, this is also called
\emph{word parallelism}.

\emph{Unrolling, fission and vectorization of loops} are methods to exploit
out-of-order execution and super scalar parallelism now standard in modern CPUs.
The processor's instruction scheduler analyses the machine code,
detects data dependencies and can dispatch multiple parallel operations.
However, only specific, simple data independencies can be detected and thus
inner loops must be designed with care (e.g. for radix sort
\cite{karkkainen2009engineering}).

\subsection{(Parallel) Atomic Sample Sort}\label{ss:atomic}

There is a huge amount of work on parallel sorting so that we can only discuss
the most relevant results. Besides (multiway)-mergesort, perhaps the most
practical parallel sorting algorithms are parallelizations of radix sort (e.g.
\cite{wassenberg2011engineering}) and quicksort \cite{TsiZha03} as well as
\emph{sample sort} \cite{BleEtAl91}. Sample sort is a generalization of
quicksort working with $k-1$ pivots at the same time.  For small inputs sample
sort uses some sequential base case sorter.  Larger inputs are split into $k$
\emph{buckets} $b_1,\ldots,b_k$ by determining $k-1$ splitter keys
$x_1\leq \cdots\leq x_{k-1}$ and then classifying the input elements -- element
$s$ goes to bucket $i$ if $x_{i-1}< s \leq x_i$ (where $x_0$ and $x_k$ are
defined as sentinel elements -- $x_0$ being smaller than all possible input
elements and $x_k$ being larger).  Splitters can be determined by drawing a
random sample of size $\alpha k-1$ from the input, sorting it, and then taking every
$\alpha$-th element as a splitter. Parameter $\alpha$ is the \emph{oversampling}
factor. The buckets are then sorted recursively and
concatenated. ``Traditional'' parallel sample sort chooses $k=p$ and uses a
sample big enough to assure that all buckets have approximately equal size.
Sample sort is also attractive as a sequential algorithm since it is more cache
efficient than quicksort and since it is particularly easy to avoid branch
mispredictions (super scalar sample sort -- S$^4$) \cite{SW04}. In this case, $k$
is chosen in such a way that classification and data distribution can be done
in a cache efficient way.

\subsection{More Related Work}

There is some work on PRAM algorithms for string sorting
(e.g. \cite{Hagerup94}). By combining pairs of adjacent characters into single
characters, one obtains algorithms with work $\Oh{N\log N}$ and time $\Oh{\log
  N/\log\log N}$. Compared to the sequential algorithms this is suboptimal
unless $D=\Oh{N}=\Oh{n}$ and with this approach it is unclear how to avoid work
on characters outside distinguishing prefixes.

We found no publications on practical parallel string sorting. However, Ta\-kuya
Akiba has implemented a parallel radix sort \cite{akiba11radixsort}, Tommi
Rantala's library \cite{rantala07web} contains multiple parallel mergesorts and
a parallel SIMD variant of multikey quicksort, and Nagaraja Shamsundar
\cite{shamsundar09lcpmergesort} also parallelized Waihong Ng's LCP-mergesort
\cite{ng2008merging}. Of all these implementations, only the radix sort by Akiba
scales fairly well to many-core architectures. For the abstract, we exclude the
other implementations and discuss their scalability issues in
Appendix~\ref{sec:more-exp-para}.


\section{Shared Memory Parallel String Sorting}\label{sec:smp-parallelize}

Already in a sequential setting, theoretical considerations and experiments (see Appendix~\ref{sec:exp-sequential})
indicate that \emph{the} best string string sorting algorithm does not exist.
Rather, it depends at least on $n$, $D$, $\sigma$, and the hardware.
Therefore we decided to parallelize several algorithms taking care that
components like data distribution, load balancing or base case sorter can be reused.
Remarkably, most algorithms in Section~\ref{sec:basic-sequential} can be
parallelized rather easily and we will discuss parallel versions
in Sections~\ref{sec:para-radixsort}--\ref{sec:para-burstsort}. However, none of
these parallelizations make use of the striking new feature of modern many-core
systems: many multi-core processors with individual cache levels but relatively
few and slow memory channels to shared RAM. Therefore we decided to design a
new string sorting algorithm based on sample sort, which exploits these
properties.
Preliminary result on string sample sort have been reported in the bachelor thesis of
Knöpfle \cite{Kno12}.

\subsection{String Sample Sort}\label{sec:s5}

In order to adapt the atomic sample sort from Section~\ref{ss:atomic} to
strings, we have to devise an efficient classification algorithm. Also, in order
to approach total work $\Oh{D+n\log n}$ we have to use the
information gained during classification in the recursive calls. This can be
done by observing that
\begin{equation}\label{eq:lcp}
\forall 1\leq i \leq k\gilt \forall s,t\in b_i\gilt \lcp(s,t)\geq \lcp_x(i)\punkt
\end{equation}
Another issue is that we have to reconcile the parallelization and load
balancing perspective from traditional parallel sample sort with the cache
efficiency perspective of super scalar sample sort. We do this by using dynamic
load balancing which includes parallel execution of recursive calls as in
parallel quicksort.

In Appendix~\ref{ss:theory} we outline a variant of string sample sort that uses
a trie data structure and a number of further tricks to enable good asymptotic
performance. However, we view this approach as somewhat risky for a first
reasonable implementation. Hence, in the following, we present a more pragmatic
implementation.

\subsubsection{Super Scalar String Sample Sort (\texorpdfstring{S$^5$}{S5}) -- A Pragmatic Solution.}

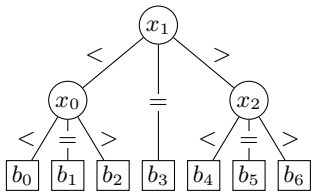
\begin{figure}[t]\centering
  \begin{tikzpicture}[
  xscale=0.6,
  bucket/.style={draw,rectangle,inner sep=2pt},
  keynode/.style={draw,circle,inner sep=1pt},
  equall/.style={fill=white,inner sep=2pt,text depth=-1pt},
  ]

\node (k1) [keynode] at (3,2) {\vphantom{0}$x_1$};

\node (k2) [keynode] at (1,1) {\vphantom{0}$x_0$};
\node (k3) [keynode] at (5,1) {\vphantom{0}$x_2$};

\node (b0)  [bucket] at (0,0) {$b_0$};
\node (b1)  [bucket] at (1,0) {$b_1$};
\node (b2)  [bucket] at (2,0) {$b_2$};
\node (b3)  [bucket] at (3,0) {$b_3$};
\node (b4)  [bucket] at (4,0) {$b_4$};
\node (b5)  [bucket] at (5,0) {$b_5$};
\node (b6)  [bucket] at (6,0) {$b_6$};

\draw (k1) -- node[left,yshift=1mm] {$<$} (k2);
\draw (k1) -- node[equall] {$=$} (b3);
\draw (k1) -- node[right,yshift=1mm] {$>$} (k3);

\draw (k2) -- node[left] {$<$} (b0);
\draw (k2) -- node[equall] {$=$} (b1);
\draw (k2) -- node[right] {$>$} (b2);

\draw (k3) -- node[left] {$<$} (b4);
\draw (k3) -- node[equall] {$=$} (b5);
\draw (k3) -- node[right] {$>$} (b6);

\end{tikzpicture}
  \caption{Ternary search tree for $v = 3$ splitters.}\label{fig:ternary-tree}
\end{figure}

We adapt the implicit binary search tree approach used in S$^4$ \cite{SW04} to strings.
Rather than using arbitrarily long splitters as in trie sample sort, or all
characters of the alphabet as in radix sort, we design the splitter keys to
consist of \emph{as many characters as fit into a machine word}. In the
following let $w$ denote the number of characters fitting into one machine word
(for 8-bit characters and 64-bit machine words we would have $w=8$).  We choose
$v=2^d-1$ splitters $x_0,\ldots,x_{v-1}$ from a sorted sample to construct a
perfect binary search tree, which is used to classify a set of strings based on
the next $w$ characters at common prefix $\ell$. The main disadvantage compared
to trie sample sort is that we may have many input strings whose next $w$
characters are identical.  For these strings, the classification does not reveal
much information. We make the best out of such inputs by explicitly defining
\emph{equality buckets} for strings whose next $w$ characters exactly match $x_i$.
For equality buckets, we can increase the common prefix length by $w$ in the
recursive calls, i.e., these characters will never be inspected again.  In
total, we have $k = 2 v + 1$ different buckets $b_0,\ldots,b_{2v}$ for a ternary
search tree (see Figure~\ref{fig:ternary-tree}).  Testing for equality can
either be implemented by explicit equality tests at each node of the search tree
(which saves time when most elements end up in a few large equality buckets) or by
going down the search tree all the way to a bucket $b_i$ ($i$ even) doing only
$\leq$-comparisons, followed by a single equality test with $x_{\frac{i}{2}}$,
unless $i = 2v$. This allows us to completely unroll the loop descending the
search tree. We can then also unroll the loop over the elements, interleaving
independent tree descents. Like in \cite{SW04}, this is an important optimization since
it allows the instruction scheduler in a super scalar processor to parallelize
the operations by drawing data dependencies apart. The strings in
the ``$< x_0$'' and ``$> x_{v-1}$'' buckets $b_0$ and $b_{2v}$ keep
common prefix length $\ell$. For other even buckets $b_i$ the common prefix
length is increased by $\lcp_x(\frac{i}{2})$.

An analysis similar to the one of multikey quicksort yields the following
asymptotic running time bound.

\begin{lemma}
  String sample sort with implicit binary trees and word parallelism
  can be implemented to run in time
  $\Oh{\frac{D}{w}\log v + n\log n}$.
\end{lemma}

\subsubsection{Implementation Details.}

Goal of S$^5$ is to have a common classification data structure that fits into
the cache of all cores. Using this data structure, all PEs can
independently classify a subset of the strings into buckets in parallel.  As
most commonly done in radix sort, we first classify strings, counting how many
fall into each bucket, then calculate a prefix sum and redistribute the string
pointers accordingly. To avoid traversing the tree twice, the bucket index of
each string is stored in an oracle. Additionally, to make higher use of super scalar parallelism, we
even separate the classification loop from the counting loop
\cite{karkkainen2009engineering}.

Like in S$^4$, the binary tree of splitters is stored in level-order as an array,
allowing efficient traversal using $i := 2 i + \set{0,1}$, without branch
mispredictions. We noticed that predicated instructions \texttt{CMOVA} were
more efficient than flag arithmetic involving \texttt{SETA}.

To perform the equality check after traversal without extra indirections, the
splitters are additionally stored in order. Another idea is to keep track of the
last $\leq$-branch during traversal; this however was slower and requires an
extra register. A third variant is to check for equality after each comparison,
which requires only an additional \texttt{JE} instruction and no extra
\texttt{CMP}. The branch misprediction cost is counter-balanced by skipping the
rest of the tree. An interesting observation is that, when breaking the tree
traversal at array index~$i$, then the corresponding equality bucket $b_j$
can be calculated from $i$ using only bit operations (note that $i$ is an index
in level-order, while $j$ is in-order). Thus in this third variant, no
additional in-order splitter array is needed.

The sample is drawn \pFORREPORT{pseudo-}randomly with an oversampling factor
$\alpha = 2$ to keep it in cache when sorting with STL's introsort and building
the search tree.
\pFORJOURNAL{Note that bad sampling does not have such an effect as in atomic
  ss? PS: no. this is more related to whether we use dynamic load balancing or
  not}
Instead of using the straight-forward equidistant method to draw splitters from
the sample, we use a simple recursive scheme that tries to avoid using the same
splitter multiple times: Select the middle sample $m$ of a range $a..b$
(initially the whole sample) as the middle splitter $\bar{x}$. Find new
boundaries $b'$ and $a'$ by scanning left and right from $m$ skipping samples
equal to $\bar{x}$. Recurse on $a..b'$ and $a'..b$.
\begin{FORREPORT}
  The splitter tree selected by this heuristic was never slower than equidistant
  selection, but slightly faster for inputs with many equal common prefixes. It
  is used in all our experiments.
\end{FORREPORT}
\pFORREPORT{The LCP of two consecutive splitters is calculated without a loop
  using two instructions: \texttt{XOR} and \texttt{BSR} to count the number of
  leading zero bits in the result.}

For current 64-bit machines with 256\,KiB L2 cache, we use $v = 8191$. Note that
the limiting data structure which must fit into L2 cache is not the splitter
tree, which is only 64\,KiB for this $v$, but is the bucket counter array
containing $2v+1$ counters, each 8 bytes long. We did not look into methods to
reduce this array's size, because the search tree must also be stored both in
level-order and in in-order.

\begin{FORJOURNAL}
Bits in binary search tree correspond to binary search intervals -> shifting around bits is exactly that.
other pointer redistribution are not commonly used (except by Jan).
\end{FORJOURNAL}

\subsubsection{Parallelization of S$^5$.}\label{sec:parallel-s5}

Parallel S$^5$ (pS$^5$) is composed of four sub-algorithms for differently
sized subsets of strings. For string sets $\Strings$ with $|\Strings| \geq
\frac{n}{p}$, a \emph{fully parallel version} of S$^5$ is run, for large sizes
$\frac{n}{p} > |\Strings| \geq t_m$ a sequential version of S$^5$ is used,
for sizes $t_m > |\Strings| \geq t_i$ the fastest sequential algorithm for
medium-size inputs (caching multikey quicksort from Section~\ref{sec:para-mkqs})
is called, which internally uses insertion sort when $|\Strings| < t_i$.

The fully parallel version of S$^5$ uses $p' = \lceil \frac{|\Strings|}{p}
\rceil$ threads for a subset $\Strings$. It consists of four stages:
selecting samples and generating a splitter tree, parallel classification and
counting, global prefix sum, and redistribution into buckets. Selecting the
sample and constructing the search tree are done sequentially, as these steps
have negligible run time. Classification is done independently, dividing the
string set evenly among the $p'$ threads. The prefix sum is done sequentially
once all threads finish counting.

In the sequential version of S$^5$ we permute the string pointer array in-place
by walking cycles of the permutation \cite{mcilroy1993engineering}. Compared to
out-of-place redistribution into buckets, the in-place algorithm uses less
input/output streams and requires no extra space. The more complex instruction
set seems to have only little negative impact, as today, memory access is the main
bottleneck. However, for fully parallel S$^5$, an in-place permutation cannot be
done in this manner. We therefore resort to out-of-place redistribution, using
an extra string pointer array of size $n$. The string pointers are not copied
back immediately. Instead, the role of the extra array and original array are
swapped for the recursion.

All work in parallel S$^5$ is dynamically load balanced via a central job
queue. Dynamic load balancing is very important and probably unavoidable for
parallel string sorting, because any algorithm must adapt to the input string
set's characteristics. We use the lock-free queue implementation from Intel's
Thread Building Blocks (TBB) and threads initiated by OpenMP to create a
light-weight thread pool.

To make work balancing most efficient, we modified all sequential sub-al\-go\-rithms
of parallel S$^5$ to use an explicit recursion stack. The traditional way to
implement dynamic load balancing would be to use work stealing among the
sequentially working threads. This would require the operations on the local
recursion stacks to be synchronized or atomic. However, for our application
fast stack operations are crucial for performance as they are very frequent. We
therefore choose a different method: voluntary work sharing. If the global job
queue is empty and a thread is idle, then a global atomic boolean flag is set to
indicate that other threads should share their work. These then free the
\emph{bottom level} of their local recursion stack (containing the largest subproblems) and enqueue this level as
separate, independent jobs. This method avoids costly atomic operations on the
local stack, replacing it by a faster (not necessarily synchronized) boolean
flag check. The short wait of an idle thread for new work does not occur often, because the largest recursive subproblems are shared. Furthermore, the global job queue never gets large because most subproblems are kept on local stacks.

\subsection{Parallel Radix Sort}\label{sec:para-radixsort}

Radix sort is very similar to sample sort, except that classification is much
faster and easier. Hence, we can use the same parallelization toolkit as with
S$^5$. Again, we use three sub-algorithms for differently sized subproblems:
fully parallel radix sort for the original string set and large subsets, a
sequential radix sort for medium-sized subsets and insertion sort for base
cases. Fully parallel radix sort consists of a counting phase, global prefix sum
and a redistribution step. Like in S$^5$, the redistribution is done
out-of-place by copying pointers into a shadow array. We experimented with 8-bit
and 16-bit radixes for the full parallel step. Smaller recursive subproblems are
processed independently by sequential radix sort (with in-place permuting), and
here we found 8-bit radixes to be faster than 16-bit sorting.  Our parallel
radix sort implementation uses the same work balancing method as parallel S$^5$.

\subsection{Parallel Caching Multikey Quicksort}\label{sec:para-mkqs}

Our preliminary experiments with sequential string sorting algorithms (see
Appendix~\ref{sec:exp-sequential}) showed a surprise winner: an enhanced variant
of multikey quicksort by Tommi Rantala \cite{rantala07web} often outperformed
more complex algorithms. This variant employs both caching of characters and
uses a super-alphabet of $w = 8$ characters, exactly as many as fit into a
machine word. The string pointer array is augmented with $w$ cache bytes for
each string, and a string subset is partitioned by a whole machine word
as splitter. Key to the algorithm's good performance, is that the cached
characters are reused for the recursive subproblems $\Strings_<$ and
$\Strings_>$, which greatly reduces the number of string accesses to at most
$\lceil \frac{D}{w} \rceil + n$ in total.

In light of this variant's good performance, we designed a parallelized
version. We use three sub-algorithms:
\emph{fully parallel caching multikey quicksort}, the original sequential caching
variant (with explicit recursion stack) for medium and small subproblems, and
insertion sort as base case. For the fully parallel sub-algorithm, we
generalized a
block-wise processing technique from (two-way) parallel atomic quicksort
\cite{TsiZha03} to three-way partitioning. 
The input array is viewed as a sequence of blocks containing
$B$ string pointers together with their $w$ cache characters. Each thread holds
exactly three blocks and performs ternary partitioning by a globally selected
pivot. When all items in a block are classified as $<$, $=$ or $>$, then the
block is added to the corresponding output set $\Strings_<$, $\Strings_=$, or
$\Strings_>$. This continues as long as unpartitioned blocks are available. If
no more input blocks are available, an extra empty memory block is allocated and
a second phase starts. The second partitioning phase ends with fully classified
blocks, which might be only partially filled. Per fully parallel partitioning
step there can be at most $3 p'$ partially filled blocks. The output sets
$\Strings_<$, $\Strings_=$, and $\Strings_>$ are processed recursively with
threads divided as evenly among them as possible. The cached characters are updated only for the $\Strings_=$ set.

In our implementation we use atomic compare-and-swap operations for block-wise
processing of the initial string pointer array and Intel TBB's lock-free queue
for sets of blocks, both as output sets and input sets for recursive steps. When
a partition reaches the threshold for sequential processing, then a continuous
array of string pointers plus cache characters is allocated and the block set is
copied into it. On this continuous array, the usual ternary partitioning scheme
of multikey quicksort is applied sequentially. Like in the other parallelized
algorithms, we use dynamic load balancing and free the bottom level when
re-balancing is required. We empirically determined $B = 128\,\text{Ki}$ as a
good block size.

\subsection{Burstsort and LCP-Mergesort}\label{sec:para-burstsort}

Burstsort is one of the fastest string sorting algorithms and cache-efficient
for many inputs, but it looks difficult to parallelize it. Keeping a common
burst trie would require prohibitively many synchronized operations, while
building independent burst tries on each PE would lead to the question how to
merge multiple tries of different structure.

One would like to generalize LCP-mergesort to a parallel $p$-way LCP-aware
merging algorithm. This looks promising in general but we leave this for future
work since LCP-mergesort is not really the best sequential algorithm in our
experiments.


\section{Experimental Results}\label{sec:experiments}

We implemented parallel S$^5$, multikey quicksort and radixsort in C++ and
compare them with Akiba's radix sort \cite{akiba11radixsort}. We also integrated
many sequential implementations into our test framework, and compiled all
programs using gcc~4.6.3 with optimizations \texttt{-O3 -march=native}. In
Appendix~\ref{sec:exp-sequential} we discuss the performance of sequential
string sorters. Our implementations and test framework are available from
\url{http://tbingmann.de/2013/parallel-string-sorting}.

We used several platforms for experiments, and summarized their properties in
Table~\ref{tab:hardware} in the appendix. Results we report in the abstract stem
from the largest machine, IntelE5, which has four 8-core Intel Xeon E5-4640
processors containing a total of 32 cores and supporting $p=64$ hardware
threads, and from a consumer-grade Intel i7 920 with four cores and $p=8$
hardware threads. Turbo-mode was disabled on IntelE5. We selected the following
datasets, all with 8-bit alphabets. More characteristics of these instances
are shown in Table~\ref{tab:data}.

\textbf{URLs} contains all URLs found on a set of web pages which were crawled breadth-first from the author's institute website. They include the protocol name.

\textbf{Random} from \cite{sinha2004cache-conscious} are strings of length $[0,20)$ over the ASCII alphabet $[33,127)$, with both length and characters chosen uniform at random.

\textbf{GOV2} is a TREC test collection consisting of 25 million HTML pages, PDF and Word documents retrieved from websites under the .gov top-level domain. We consider the whole concatenated corpus for line-based string sorting.

\textbf{Wikipedia} is an XML dump of the most recent version of all pages in the English Wikipedia, which was obtained from \url{http://dumps.wikimedia.org/}; our dump is dated \texttt{enwiki-20120601}.
Since the XML data is not line-based, we perform \emph{suffix sorting} on this input.

We also include the three largest inputs Ranjan \textbf{Sinha} \cite{sinha2004cache-conscious} tested burstsort on: a set of \textbf{URLs} excluding the protocol
name, a sequence of genomic strings of length 9 over a \textbf{DNA} alphabet,
and a list of non-duplicate English words called \textbf{NoDup}. The ``largest''
among these is NoDup with only 382\,MiB, which is why we consider these inputs
more as reference datasets than as our target.

The test framework sets up a separate run environment for each test run. The
program's memory is locked into RAM, and to isolate heap fragmentation, it was
very important to fork() a child process for each run. We use the largest prefix
$[0,2^d)$ of our inputs which can be processed with the available RAM. We
determined $t_m = 64\,\text{Ki}$ and $t_i = 64$ as good thresholds to switch
sub-algorithms.

\begin{table}[tb]\centering
\caption{Characteristics of the selected input instances.}\label{tab:data}
\def\tabcolsep{6pt}
\begin{tabular}{l|rrrrr}
Name        & $n$      & $N$                       & $\frac{D}{N}$ ($D$) & $\sigma$ & avg.\ $|s|$             \\ \hline
URLs        & 1.11\,G  & 70.7\,Gi                  & 93.5\,\%            & 84       & 68.4                    \\
Random      & $\infty$ & $\infty$                  & $-$                 & 94       & 10.5                    \\
GOV2        & 11.3\,G  & 425\,Gi                   & 84.7\,\%            & 255      & 40.3                    \\
Wikipedia   & 83.3\,G  & $\frac{1}{2} n (n\!+\!1)$ & (79.56\,T)          & 213      & $\frac{1}{2} (n\!+\!1)$ \\
Sinha URLs  & 10\,M    & 304\,Mi                   & 97.5\,\%            & 114      & 31.9                    \\
Sinha DNA   & 31.6\,M  & 302\,Mi                   & 100\,\%             & 4        & 10.0                    \\
Sinha NoDup & 31.6\,M  & 382\,Mi                   & 73.4\,\%            & 62       & 12.7                    \\
\end{tabular}
\end{table}

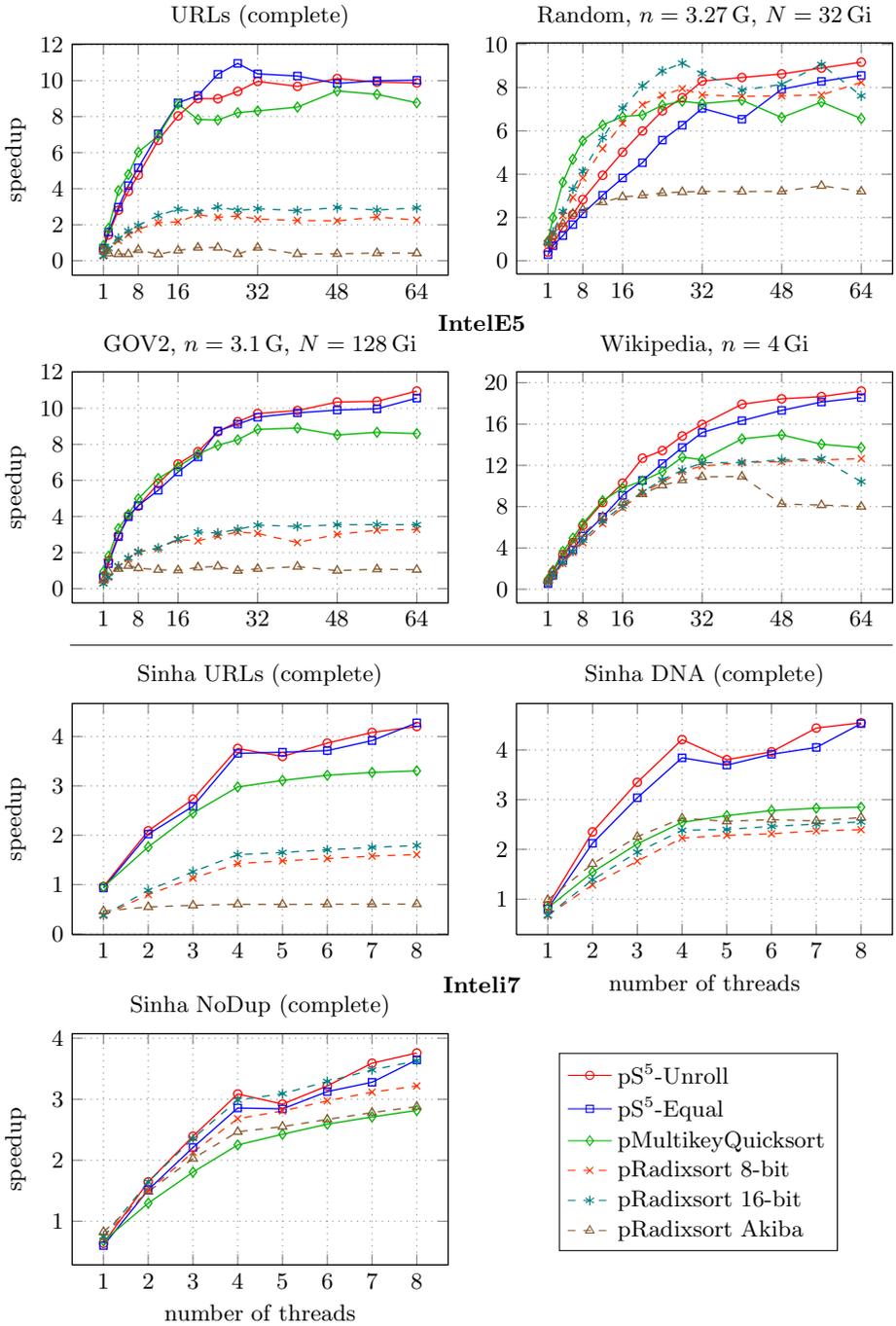
\begin{figure}[p]\centering\parskip=\smallskipamount

\pgfplotsset{
  plotSpeedup64/.append style={
    width=67mm,height=47.4mm,
  },
  plotSpeedup8/.append style={
    width=67mm,height=47.4mm,
  },
}

\begin{tikzpicture}
  \begin{axis}[plotSpeedup64,
    title={URLs (complete)},
    ylabel={speedup},
    ]

    \addplot coordinates { (1,0.640505236656836) (2,1.43438798065115) (4,2.8042754402263) (6,3.85261135170792) (8,4.75812478968528) (12,6.68392435311665) (16,8.03892320835197) (20,9.00154963437005) (24,8.99442238545133) (28,9.40461938067534) (32,9.9582813626482) (40,9.6825382523656) (48,10.1177068720401) (56,9.91866677004072) (64,9.87230508611598) };
    \addlegendentry{algo=bingmann/parallel\_sample\_sortBTCU2};
    \addplot coordinates { (1,0.694393325338075) (2,1.59158054378123) (4,2.97833798944614) (6,4.17417109474535) (8,5.15767244523796) (12,7.04605567700049) (16,8.76826872238069) (20,9.17456945862602) (24,10.3522711509117) (28,10.958453747647) (32,10.3769122990656) (40,10.2549538846185) (48,9.84181759568607) (56,9.99441980893426) (64,10.0236919965824) };
    \addlegendentry{algo=bingmann/parallel\_sample\_sortBTCEU1};
    \addplot coordinates { (1,0.843028473525333) (2,1.80161681324835) (4,3.89483603467772) (6,4.7886565780796) (8,6.03120012032857) (12,6.94037244135476) (16,8.7077347923041) (20,7.83925697786895) (24,7.81031372600322) (28,8.2193348242831) (32,8.31628491520676) (40,8.52921801162276) (48,9.4284821876656) (56,9.23793198897602) (64,8.76990020924395) };
    \addlegendentry{algo=bingmann/parallel\_mkqs};
    \addplot coordinates { (1,0.26638044717664) (2,0.609654562061395) (4,1.11883726014847) (6,1.45327375800589) (8,1.73782379062645) (12,2.11231537938382) (16,2.15896205756257) (20,2.55773414256813) (24,2.41082934008148) (28,2.48420429694905) (32,2.31353418098679) (40,2.23657077684492) (48,2.21319888952047) (56,2.42975490639414) (64,2.2543718323096) };
    \addlegendentry{algo=bingmann/parallel\_radix\_sort\_8bit};
    \addplot coordinates { (1,0.264588626918001) (2,0.682434286062651) (4,1.25612732492355) (6,1.65103966617605) (8,1.96214591679113) (12,2.5106699125754) (16,2.86244885782475) (20,2.72439459700645) (24,2.97807387043852) (28,2.80779239781361) (32,2.89090948220204) (40,2.77994443380949) (48,2.95402241977426) (56,2.81382098493894) (64,2.93594444110528) };
    \addlegendentry{algo=bingmann/parallel\_radix\_sort\_16bit};
    \addplot coordinates { (1,0.604591971644772) (2,0.406450978463959) (4,0.370447792064532) (6,0.371418301466827) (8,0.601292970800167) (12,0.354727245119592) (16,0.568779758691284) (20,0.729358202897528) (24,0.728889226515964) (28,0.372223646200633) (32,0.726901311846664) (40,0.372411976711949) (48,0.38310754926846) (56,0.425334051041733) (64,0.418910848633688) };
    \addlegendentry{algo=akiba/parallel\_radix\_sort};

    \legend{}
  \end{axis}
\end{tikzpicture}
\hfill%
\begin{tikzpicture}
  \begin{axis}[plotSpeedup64,
    title={Random, $n = 3.27\,\text{G}$, $N = 32\,\text{Gi}$},
    ]

    \addplot coordinates { (1,0.40419670621986) (2,0.796823001846478) (4,1.54460963881543) (6,2.13660978011159) (8,2.83240131270511) (12,3.95186214330942) (16,5.01956809284209) (20,5.99589522909703) (24,6.92198900091659) (28,7.5308037495014) (32,8.30069576495675) (40,8.46510559117607) (48,8.6288084369769) (56,8.90243370525584) (64,9.17344278783432) };
    \addlegendentry{algo=bingmann/parallel\_sample\_sortBTCU2};
    \addplot coordinates { (1,0.282920734585615) (2,0.69801462228836) (4,1.16592971589313) (6,1.67629797340792) (8,2.17621801688092) (12,3.03111027269151) (16,3.82709411281781) (20,4.5300649046825) (24,5.57330942207069) (28,6.26188008391307) (32,7.03961706609991) (40,6.5398484520459) (48,7.92197288531015) (56,8.28686898392201) (64,8.55981701311081) };
    \addlegendentry{algo=bingmann/parallel\_sample\_sortBTCEU1};
    \addplot coordinates { (1,0.914383097227267) (2,1.99091788662253) (4,3.63108294587434) (6,4.69070231122319) (8,5.54377014160604) (12,6.2693658317906) (16,6.6599848313814) (20,6.73199975039892) (24,7.18399748860837) (28,7.36884782014754) (32,7.26192147548393) (40,7.41653817824699) (48,6.61402172009108) (56,7.33036632952185) (64,6.56423517549502) };
    \addlegendentry{algo=bingmann/parallel\_mkqs};
    \addplot coordinates { (1,0.789366187834678) (2,1.23659773472693) (4,2.08436099273554) (6,2.9298372892403) (8,3.82614400940337) (12,5.17515041870537) (16,6.35386815868075) (20,7.21212671066077) (24,7.63878113541535) (28,7.95078903676346) (32,7.66128938172636) (40,7.59181535696226) (48,7.61820648747346) (56,7.6607531323878) (64,8.24129942041606) };
    \addlegendentry{algo=bingmann/parallel\_radix\_sort\_8bit};
    \addplot coordinates { (1,0.81574765895807) (2,1.33631259411568) (4,2.28840990769865) (6,3.30876405872791) (8,4.13107266133135) (12,5.68452389913436) (16,7.04506782096013) (20,8.07689633892657) (24,8.76624250993065) (28,9.12858902498431) (32,8.6473937209297) (40,7.8763072180023) (48,8.14979128805701) (56,9.06585082424574) (64,7.61569428673426) };
    \addlegendentry{algo=bingmann/parallel\_radix\_sort\_16bit};
    \addplot coordinates { (1,0.877229114394573) (2,1.15610738850948) (4,1.71465126364282) (6,2.17289936095342) (8,2.36751949187878) (12,2.71191255103763) (16,2.94984180305457) (20,3.02443371139305) (24,3.13653165650491) (28,3.1788330078125) (32,3.20944573356793) (40,3.19945178086486) (48,3.20736021745132) (56,3.46633220724856) (64,3.20646141957617) };
    \addlegendentry{algo=akiba/parallel\_radix\_sort};

    \legend{}
  \end{axis}
\end{tikzpicture}

\hfill%
\begin{tikzpicture}
  \path (0,0) -- node[overlay,below,outer sep=-4pt] {\textbf{IntelE5}} (112mm,0);
\end{tikzpicture}

\begin{tikzpicture}
  \begin{axis}[plotSpeedup64,
    title={GOV2, $n = 3.1\,\text{G}$, $N = 128\,\text{Gi}$},
    ylabel={speedup},
    ]

    \addplot coordinates { (1,0.69015370748138) (2,1.54532321370717) (4,2.94969535376474) (6,4.04625402862715) (8,4.57097297445908) (12,5.8227392120075) (16,6.90858364003324) (20,7.59054296429154) (24,8.69872638124851) (28,9.25164552132023) (32,9.70869412883528) (40,9.87426186112808) (48,10.3446988551518) (56,10.3830918563985) (64,10.9511263338164) };
    \addlegendentry{algo=bingmann/parallel\_sample\_sortBTCU2};
    \addplot coordinates { (1,0.627585686244904) (2,1.38987977591291) (4,2.87901801396193) (6,3.98185307974176) (8,4.60588647345863) (12,5.44821290639702) (16,6.46469696212235) (20,7.30159825587779) (24,8.74188302542975) (28,9.12320584477177) (32,9.50460603684829) (40,9.74274831526516) (48,9.90004848711689) (56,9.96899198245734) (64,10.5612675481184) };
    \addlegendentry{algo=bingmann/parallel\_sample\_sortBTCEU1};
    \addplot coordinates { (1,0.941244025330174) (2,1.79763583594499) (4,3.3285570875769) (6,4.11239456825546) (8,4.98138113459514) (12,6.10528524590164) (16,6.75834035016405) (20,7.45617856593716) (24,7.94655049980882) (28,8.24918664899385) (32,8.82751820388349) (40,8.90213026351015) (48,8.51804142574837) (56,8.66472598405265) (64,8.59294590184511) };
    \addlegendentry{algo=bingmann/parallel\_mkqs};
    \addplot coordinates { (1,0.333232164740646) (2,0.628141191709845) (4,1.21301883075666) (6,1.65915606605147) (8,2.02739850116541) (12,2.21840570317563) (16,2.72494199237179) (20,2.64663945330711) (24,2.92746609248601) (28,3.13613581244947) (32,3.06003486455148) (40,2.55502159151531) (48,3.01022704346881) (56,3.23648336735318) (64,3.28507515348246) };
    \addlegendentry{algo=bingmann/parallel\_radix\_sort\_8bit};
    \addplot coordinates { (1,0.277557970651575) (2,0.64423667735025) (4,1.25417989919748) (6,1.70265869433866) (8,2.045496487494) (12,2.25818281303662) (16,2.75201763077061) (20,3.1380045783126) (24,3.0840371095665) (28,3.28762711864407) (32,3.52030538230259) (40,3.44889716014663) (48,3.54132181110029) (56,3.54639363744401) (64,3.54666382239559) };
    \addlegendentry{algo=bingmann/parallel\_radix\_sort\_16bit};
    \addplot coordinates { (1,0.452235882928952) (2,0.951727521772908) (4,1.10300054969009) (6,1.2728789998261) (8,1.14064215148189) (12,1.04749561225867) (16,1.02159372202033) (20,1.18243267270099) (24,1.23175367889862) (28,0.998138920573933) (32,1.10467566490119) (40,1.22379449691007) (48,1.00213546420514) (56,1.08042220963804) (64,1.05198362122732) };
    \addlegendentry{algo=akiba/parallel\_radix\_sort};

    \legend{}
  \end{axis}
\end{tikzpicture}
\hfill%
\begin{tikzpicture}
  \begin{axis}[plotSpeedup64,
    title={Wikipedia, $n = 4\,\text{Gi}$},
    ytick={0,4,8,12,16,20},
    ]

    \addplot coordinates { (1,0.726163354185809) (2,1.65000214503282) (4,3.3563522652898) (6,4.51080465611775) (8,6.18132044743496) (12,8.39921818982555) (16,10.267812251109) (20,12.6897697028353) (24,13.42622922206) (28,14.8390166738472) (32,15.963778743186) (40,17.9196757210082) (48,18.4421002157756) (56,18.6598204900137) (64,19.1932331389885) };
    \addlegendentry{algo=bingmann/parallel\_sample\_sortBTCU2};
    \addplot coordinates { (1,0.54020787490051) (2,1.33622466705269) (4,2.79488368882112) (6,3.82913056969198) (8,5.14631698668629) (12,6.99978160375642) (16,9.09032179280624) (20,10.5343741440701) (24,12.1672237981251) (28,13.7221042867082) (32,15.1707952035342) (40,16.3266663365006) (48,17.3236042609734) (56,18.1352802030696) (64,18.5637519105462) };
    \addlegendentry{algo=bingmann/parallel\_sample\_sortBTCEU1};
    \addplot coordinates { (1,0.972465233881163) (2,1.80658545750611) (4,3.66177828696196) (6,4.95046658800815) (8,6.36272898721752) (12,8.57241563767664) (16,9.7965681488211) (20,10.4917958253959) (24,11.3828669517735) (28,12.7855282841155) (32,12.5476997025735) (40,14.5639633953929) (48,14.9506323209288) (56,14.0433533750396) (64,13.7052346506076) };
    \addlegendentry{algo=bingmann/parallel\_mkqs};
    \addplot coordinates { (1,0.751372400903863) (2,1.28112942534989) (4,2.48246794530451) (6,3.57987096293791) (8,4.51084874310958) (12,6.34453596683199) (16,7.88959773533816) (20,9.31879580834696) (24,10.394349829513) (28,11.4159777978955) (32,11.9278023869457) (40,12.2623943886498) (48,12.3572786419984) (56,12.5170724827919) (64,12.6491481944353) };
    \addlegendentry{algo=bingmann/parallel\_radix\_sort\_8bit};
    \addplot coordinates { (1,0.649583957303548) (2,1.32360937446228) (4,2.61261781438397) (6,3.65063294343216) (8,4.75369042076771) (12,6.62057212695698) (16,8.07320127902827) (20,9.4338450217689) (24,10.6082239639598) (28,11.5180282702444) (32,12.2406046943376) (40,12.2775954074602) (48,12.5240695112288) (56,12.6364034607381) (64,10.4096352933216) };
    \addlegendentry{algo=bingmann/parallel\_radix\_sort\_16bit};
    \addplot coordinates { (1,0.863826251010691) (2,1.7682405406648) (4,3.17478617251204) (6,4.36854938248234) (8,5.37436274840876) (12,7.02276648904281) (16,8.37936368455835) (20,9.22245046398798) (24,10.0627488259123) (28,10.5582804121447) (32,10.8769283704357) (40,10.9156701733142) (48,8.24894906917555) (56,8.14894856719106) (64,8.00284371694612) };
    \addlegendentry{algo=akiba/parallel\_radix\_sort};

    \legend{}
  \end{axis}
\end{tikzpicture}

\hfill%
\begin{tikzpicture}
  \draw (0,0) -- (112mm,0);
\end{tikzpicture}

\begin{tikzpicture}
  \begin{axis}[plotSpeedup8,
    title={Sinha URLs (complete)},
    ylabel={speedup},
    ]

    \addplot coordinates { (1,0.95938473615779) (2,2.08810900109031) (3,2.73007336294658) (4,3.75711170248537) (5,3.59599017220706) (6,3.86534156333859) (7,4.08185972332107) (8,4.20176698680723) };
    \addlegendentry{algo=bingmann/parallel\_sample\_sortBTCU2};
    \addplot coordinates { (1,0.930494948677805) (2,2.02111380025148) (3,2.58503725120419) (4,3.65688284947247) (5,3.68145528348902) (6,3.71354694374044) (7,3.91633252284262) (8,4.27708517663425) };
    \addlegendentry{algo=bingmann/parallel\_sample\_sortBTCEU1};
    \addplot coordinates { (1,0.953380936476864) (2,1.76157078958392) (3,2.44783069975307) (4,2.97710498970391) (5,3.10957384635569) (6,3.21399425535325) (7,3.27087506534239) (8,3.30401958042667) };
    \addlegendentry{algo=bingmann/parallel\_mkqs};
    \addplot coordinates { (1,0.376299358295253) (2,0.796137732230759) (3,1.12765135350227) (4,1.4256384083991) (5,1.47893204250652) (6,1.52620199812676) (7,1.57548826866056) (8,1.60724149268453) };
    \addlegendentry{algo=bingmann/parallel\_radix\_sort\_8bit};
    \addplot coordinates { (1,0.376409825998741) (2,0.882558438686207) (3,1.25853696212553) (4,1.60545340345087) (5,1.65000379724404) (6,1.70277966455928) (7,1.75231992830738) (8,1.79023021391721) };
    \addlegendentry{algo=bingmann/parallel\_radix\_sort\_16bit};
    \addplot coordinates { (1,0.464543396298135) (2,0.546757057913125) (3,0.580494618452673) (4,0.601142723629662) (5,0.598778172464478) (6,0.602253634783037) (7,0.603961298193864) (8,0.604987183195987) };
    \addlegendentry{algo=akiba/parallel\_radix\_sort};

    \legend{}
  \end{axis}
\end{tikzpicture}
\hfill%
\begin{tikzpicture}
  \begin{axis}[plotSpeedup8,
    title={Sinha DNA (complete)},
    xlabel={number of threads},
    every axis x label/.append style={overlay},
    ]

    \addplot coordinates { (1,0.869675494894935) (2,2.3511356359592) (3,3.34828215130333) (4,4.20827554512854) (5,3.80390718332829) (6,3.96257437256459) (7,4.44044397148592) (8,4.54570549821339) };
    \addlegendentry{algo=bingmann/parallel\_sample\_sortBTCU2};
    \addplot coordinates { (1,0.80665408732307) (2,2.12236352271533) (3,3.03760559254538) (4,3.84202826424152) (5,3.69489077103085) (6,3.91285082219909) (7,4.04997619728702) (8,4.52849407032519) };
    \addlegendentry{algo=bingmann/parallel\_sample\_sortBTCEU1};
    \addplot coordinates { (1,0.822294446748343) (2,1.54089867650337) (3,2.11053139671442) (4,2.54341463737997) (5,2.67755554935239) (6,2.78010006485014) (7,2.82782103676625) (8,2.84948384701077) };
    \addlegendentry{algo=bingmann/parallel\_mkqs};
    \addplot coordinates { (1,0.673308209652582) (2,1.28543262570544) (3,1.76319641187189) (4,2.22687634397681) (5,2.28197327766428) (6,2.31376864906589) (7,2.37078436461152) (8,2.39429263209131) };
    \addlegendentry{algo=bingmann/parallel\_radix\_sort\_8bit};
    \addplot coordinates { (1,0.673450617576488) (2,1.38846046674893) (3,1.94254136373762) (4,2.38270983928261) (5,2.40036660858903) (6,2.46171564224304) (7,2.50742884030362) (8,2.55173698054356) };
    \addlegendentry{algo=bingmann/parallel\_radix\_sort\_16bit};
    \addplot coordinates { (1,0.987156364571943) (2,1.70751929221814) (3,2.2547769952017) (4,2.62501710407487) (5,2.56423742803009) (6,2.59762161869396) (7,2.57182596338823) (8,2.63967252025249) };
    \addlegendentry{algo=akiba/parallel\_radix\_sort};

    \legend{}
  \end{axis}
\end{tikzpicture}

\hfill%
\begin{tikzpicture}
  \path (0,0) -- node[overlay,below,outer sep=-3pt] {\textbf{Inteli7}} (112mm,0);
\end{tikzpicture}

\begin{tikzpicture}
  \begin{axis}[plotSpeedup8,
    title={Sinha NoDup (complete)},
    legend to name={speedupFront},legend columns=1,
    ylabel={speedup},xlabel={number of threads},
    ]

    \addplot coordinates { (1,0.639551375594504) (2,1.64546158555622) (3,2.39296559450456) (4,3.08629159822352) (5,2.92060076571001) (6,3.21811600236303) (7,3.59061812937674) (8,3.75738357559485) };
    \addlegendentry{algo=bingmann/parallel\_sample\_sortBTCU2};
    \addplot coordinates { (1,0.601229020650095) (2,1.51500784876613) (3,2.20904662112511) (4,2.85608219852915) (5,2.84182258696992) (6,3.12495668712318) (7,3.27796489697675) (8,3.64454801803258) };
    \addlegendentry{algo=bingmann/parallel\_sample\_sortBTCEU1};
    \addplot coordinates { (1,0.67706534668662) (2,1.29604051932955) (3,1.80462144237204) (4,2.25047550919687) (5,2.42578474063138) (6,2.59106739833427) (7,2.70923316221286) (8,2.81624509905971) };
    \addlegendentry{algo=bingmann/parallel\_mkqs};
    \addplot coordinates { (1,0.750040659121319) (2,1.49697164187913) (3,2.13228707208852) (4,2.68108833021183) (5,2.80689559774527) (6,2.97298683222534) (7,3.115456953134) (8,3.21689158393278) };
    \addlegendentry{algo=bingmann/parallel\_radix\_sort\_8bit};
    \addplot coordinates { (1,0.74985426403569) (2,1.64309659343952) (3,2.35809369970309) (4,2.98630963550002) (5,3.09087516469791) (6,3.29041621229381) (7,3.48014183484901) (8,3.6289702410759) };
    \addlegendentry{algo=bingmann/parallel\_radix\_sort\_16bit};
    \addplot coordinates { (1,0.824336758633157) (2,1.48591105806196) (3,2.02674290850979) (4,2.46621252081333) (5,2.55073693472864) (6,2.66917688842119) (7,2.78025354438054) (8,2.87886472701735) };
    \addlegendentry{algo=akiba/parallel\_radix\_sort};

    \SpeedupLegend
  \end{axis}
\end{tikzpicture}
\hfill%
\parbox{53mm}{\hfill\ref{speedupFront}\hfill\null}
\caption{Speedup of parallel algorithm implementations on IntelE5 (top four plots) and Inteli7 (bottom three plots)}\label{fig:main-result}
\end{figure}

Figure~\ref{fig:main-result} shows a selection of 
the detailed parallel measurements from Appendix~\ref{sec:more-exp-para}.
For large instances we show results on IntelE5 (median of 1--3 repetitions) and for small instances on Inteli7 (of ten repetitions).
The plots show
the speedup of our implementations and
Akiba's radix sort over the best sequential algorithm from
Appendix~\ref{sec:exp-sequential}. We included pS$^5$-Unroll, which interleaves
three unrolled descents of the search tree, pS$^5$-Equal, which unrolls a single
descent testing equality at each node, our parallel multikey quicksort (pMKQS),
and radix sort with 8-bit and 16-bit fully parallel steps. 
On all platforms, our parallel implementations yield good speedups, limited by
memory bandwidth, not processing power. On IntelE5 for all four test instances,
pMKQS is fastest for a small number of threads. But for higher numbers, pS$^5$
becomes more efficient than pMKQS, because it utilizes memory bandwidth
better. On all instances, except Random, pS$^5$ yields the highest speedup for
both the number of physical cores and hardware threads. On Random, our 16-bit
parallel radix sort achieves a slightly higher speedup. Akiba's radix sort does
not parallelize recursive sorting steps (only the top-level is parallelized) and only performs simple load
balancing. This can be seen most pronounced on URLs and GOV2. On Inteli7, pS$^5$
is consistently faster than pMKQS for Sinha's smaller datasets, achieving
speedups of 3.8--4.5, which is higher than the three memory channels on this
platform. On IntelE5, the highest speedup of 19.2 is gained with pS$^5$ for
suffix sorting Wikipedia, again higher than the $4 \times 4$ memory channels. For all test instances, except URLs, the fully
parallel sub-algorithm of pS$^5$ was run only 1--4 times, thus most of the
speedup is gained in the sequential S$^5$ steps. The pS$^5$-Equal variant
handles URL instances better, as many equal matches occur here. However, for all
other inputs, interleaving tree descents fares better.  Overall, pS$^5$-Unroll is
currently the best parallel string sorting implementation on these platforms.

\section{Conclusions and Future Work}\label{sec:conclusions}

We have demonstrated that string sorting can be parallelized successfully on
modern multi-core shared memory machines. In particular, our new string sample
sort algorithm combines favorable features of some of the best sequential
algorithms -- robust multiway divide-and-conquer from burstsort, efficient data
distribution from radix sort, asymptotic guarantees similar to multikey
quicksort, and word parallelism from cached multikey quicksort.

Implementing some of the refinements discussed in Appendix~\ref{app:refinements}
are likely to yield further improvements for string sample sort. To improve
scalability on large machines, we may also have to look at NUMA (non uniform
memory access) effects more explicitly. Developing a parallel multiway LCP-aware
mergesort might then become interesting.

\bibliographystyle{splncs}
\bibliography{diss,library}


\clearpage
\begin{appendix}
\section{More on String Sample Sort}

\subsection{Trie Sample Sort -- A Theoretical Solution.}\label{ss:theory}

We would like to design a classification algorithm here that fulfills two
conflicting goals. On the one hand, we would like to classify the input strings
into the $k$ buckets defined by the splitters exactly, regardless how long the
substrings are that need to be inspected. The classifier from S$^5$ falls short
of this goal since it looks only at a fixed number of input characters which may
only allow a very rough classification if the splitters have long common
prefixes. On the other hand, we do not want to repeatedly inspect long common
substrings in recursive calls. In the following, we outline such an
algorithm. Since the parallelization is analogous to S$^5$, we focus on the amount of work needed by the algorithm.

Our starting point is to use a Patricia trie as a classification data structure,
adapting a technique for String B-Trees \cite{FerGro99}: When classifying $s$
using \emph{blind search}, it suffices to check the first character for each trie edge
traversed. When the search has reached a leaf, we can compare $s$ with the
splitter associated with the leaf. The first mismatch tells us in which bucket
$s$ has to be placed. When the
mismatched character of $s$ is smaller than the label character, then the output
bucket is associated with the splitter in the leftmost leaf of the current
sub-trie. If the character is larger, then the output bucket is associated with
the splitter \emph{following} the rightmost leaf of the current
sub-trie. Pointers to these buckets can be precomputed and stored for each trie
edge.  The advantage of this approach over an ordinary trie is that only a
single access to a splitter key is needed and all the other information needed
fits into a cache of size $\Oh{k}$.

However, we have the problem that the tries might be very deep (up to
$\Om{k}$) in the worst case. This will destroy the desired time bounds if the
classification frequently runs all the way down the trie and later finds a
mismatch high up in the trie (in a character ignored by the blind search). Going
beyond \cite{FerGro99}, we therefore augment the trie edges with a hash
signature of the entire substring associated with this edge -- not just its
first character. We can then stop traversing a trie as soon as there is a
mismatch in the hash signature. We will still sometimes miss the first
opportunity to stop traversing the trie, but the expected cost for finding the
first mismatch is proportional to the actual common prefix length.

We also have
to explain how to handle the case when there is a mismatch for the first
character in a trie-edge label. In this case, we have to locate the next
character of $s$ among the first characters of all edge labels. Using hashing
and van Emde Boas trees, this can be done in expected time $\Oh{\log\log\sigma}$
\cite{MehNae90} using space proportional to the degree of the current trie-node.

When a string $s$ is located into bucket $i$, we find $\lcp(s,x_i)$ equal
characters. In light of the analysis of multikey quicksort, this is unfortunate since
only $\lcp_x(i)$ of those will not be considered again.  We address this problem
by changing the splitter keys. Rather than using the complete input strings, we
only use their distinguishing prefixes. This does not get rid of all cases where
characters of $s$ will have to be reinspected later. However, this now only
happens when bucket $i+1$ has common prefix length at least $\lcp(s,x_i)$.  In
an amortized analysis, and assuming that buckets have about equal size, we can
therefore charge the comparisons of characters $s[\lcp_x(i)..\lcp(s,x_i)]$ to
characters in bucket $i+1$. We did not yet do a rigorous analysis taking the
probabilistic nature of bucket sizes into account. However, we conjecture that
an expected time bound of
$$\Oh{D+n\log_k n\cdot\log\log\sigma}$$
can be shown for trie sample sort. Choosing a small fixed $k$ we can then obtain
an algorithm that is reasonably cache efficient. Choosing $k=\sqrt{n}$ could
reduce the number of recursion levels to $\Oh{\log\log n}$ leading to expected
work $\Oh{D+n\log\log n\cdot\log\log\sigma}$.

\subsection{Practical Refinements}\label{app:refinements}

\paragraph*{Multipass data distribution:} There are two constraints for the
maximum sensible value for $k$: The cache size needed
for the classification data structure and the resources needed for data
distribution. Already in the plain external memory model these two constraints
differ ($k=\Oh{M}$ versus $k=\Oh{M/B}$). In practice, things are even more
complicated since multiple cache levels, cache replacement policy, TLBs,
etc. play a role. Anyway, we can increase the value of $k$ to the value required for
classification by doing the data distribution in multiple passes (usually two).
Note that this fits very well with our approach to compute oracles even for
single pass data distribution. This approach can be viewed as LSD radix sort
using the oracles as keys. Initial experiments indicate that this could indeed lead
to some performance improvements.

\paragraph*{Alphabet compression:} When we know that only $\sigma'<\sigma$
different values from $\Sigma$ appear in the input, we can compress characters
into $\ceil{\log\sigma'}$ bits.  For the pragmatic solution, this
allows us to pack more characters into a single machine word.  For example, for
DNA input, we might pack 32 characters into a single 64 bit machine word. Note
that this compression can be done on the fly without changing the input/output
format and the compression overhead is amortized over $\log k$ key comparisons.

\paragraph*{Jump tables:} In the pragmatic solution, the $a$ most significant
bits of a key are often already sufficient to define a path in the search tree
of length up to $a$.  We can exploit this by precomputing a jump table of size
$2^a$ storing a pointer to the end of this path. During element classification,
a lookup in this jump table can replace the traversal of the path.  This might
reduce the gap to radix sort for easy instances.

\paragraph*{Using tries in practice:} The success of burstsort indicates that
traversing tries can be made efficient. Thus, we might also be able to use a
tuned trie based implementation in practice. One ingredient to such an
implementation could be the word parallelism used in the pragmatic solution --
we define the trie over an enlarged alphabet. This reduces the number of
required hash table accesses by a factor of $w$. The tuned van Emde Boas trees
from \cite{DKMS04} suggest that this data structure might work in practice.

\paragraph*{Adaptivity:} By inspecting the sample, we can adaptively tune the
algorithm.  For example, when noticing that already a lot of information%
\footnote{The entropy $\frac{1}{n}\sum_i\log\frac{n}{|b_i|}$ can be used to
  define the amount of information gained by a set of splitters. The bucket
  sizes $b_i$ can be estimated using their size within the sample.} can be
gained from a few most significant bits in the sample keys, the algorithm might
decide to switch to radix sort. On the other hand, when even the $w$ most
significant characters do not give a lot of information, then a trie based
implementation can be used.  Again, this trie can be adapted to the input, for
example, using hash tables for low degree trie nodes and arrays for high degree
nodes.


\section{More Experimental Results}

\subsection{Experimental Setup}

We tested our implementations and those by other authors on five different test
platforms. All platforms run Linux and their main properties are listed in
Table~\ref{tab:hardware}.

As described in Section~\ref{sec:experiments}, we isolate runs of different
algorithms using following methods: before each run, the program fork()s into a
child process. The string data is loaded before the fork(), allocating exactly
the matching amount of RAM, and shared read-only with the child processes. The
program's memory is locked into RAM using mlockall(). Before the algorithm is
called, the string pointer array is generated inside the child process by
scanning the string data for NUL characters (thus flushing caches and TLB
entries). Time measurement is done with clock\_gettime() and encompasses only
the sorting algorithm. Because many algorithms have a deep recursion stack for
our large inputs, we increased the stack size limit to 64\,MiB. We took no
special precautions of pinning threads to specific cores, and used the regular
Linux task scheduling system as is. 

The output of each string sorting algorithm was verified by first checking that
the output pointer list is a permutation of the input set, and then checking
that strings are in non-descending order.

Methodologically we have to discuss, whether measuring only the algorithms run
time is a good decision. The issue is that deallocation and defragmentation in
both heap allocators and kernel page tables is done lazily. This was most
notable when running two algorithms consecutively. The fork() process isolation
was done to exclude both variables from the experimental results, however, for
use in a real program context these costs cannot not be ignored. We currently do
not know how to invoke the lazy cleanup procedures to regenerate a pristine
memory environment. These issues must be discussed in greater detail in future
work for sound results with big data in RAM.
We also did not look into HugePages (yet), which may or may not yield a
performance boost.

\begin{table}\centering
\caption{Hard- and software characteristics of experimental platforms}\label{tab:hardware}
\begin{tabular}{l|l|r|r||r|r|r|r}
Name    & Processor          & Clock & Sockets $\times$                     & Cache: L1      & L2              & L3            & RAM   \\
        &                    & [GHz] & Cores $\times$ HT                    & [KiB]          & [KiB]           & [MiB]         & [GiB] \\\hline
IntelE5 & Intel Xeon E5-4640 & 2.4   & $4 \times 8 \times 2$                & $32 \times 32$ & $32 \times 256$ & $4 \times 20$ & 512   \\
AMD48   & AMD Opteron 6168   & 1.9   & $4 \times 12 \:\phantom{\:\times 1}$ & $48 \times 64$ & $48 \times 512$ & $8 \times 6$  & 256   \\
AMD16   & AMD Opteron 8350   & 2.0   & $4 \times 4 \:\phantom{\:\times 1}$  & $16 \times 64$ & $16 \times 512$ & $4 \times 2$  & 64    \\
Inteli7 & Intel Core i7 920  & 2.67  & $1 \times 4 \times 2$                & $4 \times 32$  & $4 \times 256$  & $1 \times 8$  & 12    \\
IntelX5 & Intel Xeon X5355   & 2.67  & $2 \times 4 \times 1$                & $8 \times 32$  & $4 \times 4096$ &               & 16    \\ 
\end{tabular}

\bigskip
\begin{tabular}{l|l|l|l|l}
Name     & Codename     & Memory Channels      & Interconnect            & Linux/Kernel Version   \\ \hline
IntelE5  & Sandy Bridge & 4 $\times$ DDR3-1600 & 2 $\times$ 8.0 GT/s QPI & Ubuntu 12.04/3.2.0-38  \\
AMD48    & Magny-Cours  & 4 $\times$ DDR3-667  & 4 $\times$ 3.2 GHz HT   & Ubuntu 12.04/3.2.0-38  \\
AMD16    & Barcelona    & 2 $\times$ DDR2-533  & 3 $\times$ 1.0 GHz HT   & Ubuntu 10.04/2.6.32-45 \\
Inteli7  & Bloomfield   & 3 $\times$ DDR3-800  & 1 $\times$ 4.8 GT/s QPI & openSUSE 11.3/2.6.34   \\
IntelX5  & Clovertown   & 2 $\times$ DDR2-667  & 1 $\times$ 1.3 GHz FSB  & Ubuntu 12.04/3.2.0-38  \\
\end{tabular}
\end{table}

\subsection{Performance of Sequential Algorithms}\label{sec:exp-sequential}

We collected many sequential string sorting algorithms in our test framework.

The algorithm library by Tommi Rantala \cite{rantala07web} contains 37~versions
of radix sort (in-place, out-of-place, and one-pass with various dynamic memory
allocation schemes), 26~variants of multikey quicksort (with caching,
block-based, different dynamic memory allocation and SIMD instructions),
10~different funnelsorts, 38~implementations of burstsort (again with different
dynamic memory managements), and 29~mergesorts (with losertree and LCP caching
variants). In total these are 140~original implementation variants, all of high
quality.

The other main source of string sorting implementations are the publications of
Ranjan Sinha. We included the original burstsort implementations (one with
dynamically growing arrays and one with linked lists), and 9~versions of
copy-burstsort. The original copy-burstsort code was written for 32-bit
machines, and we modified it to work with 64-bit pointers.

We also incorporated the implementations of CRadix sort and LCP-Merge\-sort by
Waihong Ng, and the original multikey quicksort code by Bentley and Sedgewick.

\begin{table}[tb]\centering
\caption{Description of selected sequential algorithms}\label{tab:seqalgo}
\begin{tabular}{l|p{67ex}}
Name             & Description and Author                                                                                                                                        \\ \hline
mkqs             & Original multikey quicksort by Bentley and Sedgewick \cite{BenSed97}.                                                                                         \\
mkqs\_cache8     & Modified multikey quicksort with caching of eight characters by Tommi Rantala \cite{rantala07web}, slightly improved.                                         \\
radix8\_CI       & 8-bit in-place radix sort by Tommi Rantala \cite{karkkainen2009engineering}.                                                                                  \\
radix16\_CI      & Adaptive 16-/8-bit in-place radix sort by Tommi Rantala \cite{karkkainen2009engineering}.                                                                     \\
radixR\_CE7      & Adaptive 16-/8-bit out-of-place radix sort by Tommi Rantala~\cite{karkkainen2009engineering}, version CE7 (preallocated swap array, unrolling, sorted-check). \\
CRadix           & Cache efficient radix sort by Waihong Ng \cite{ng2007cache}, unmodified.                                                                                      \\
LCPMergesort     & LCP-mergesort by Waihong Ng \cite{ng2008merging}, unmodified.                                                                                                 \\
Seq-S$^5$-Unroll & Sequential Super Scalar String Sample Sort with interleaved loop over strings, unrolled tree traversal and radix sort as base sorter.                         \\
Seq-S$^5$-Equal  & Sequential Super Scalar String Sample Sort with equality check, unrolled tree traversal and radix sort as base sorter.                                        \\
burstsortA       & Burstsort using dynamic arrays by Ranjan Sinha \cite{sinha2004cache-conscious}, from \cite{rantala07web}.                                                     \\
fbC-burstsort    & Copy-Burstsort with ``free bursts'' by Ranjan Sinha \cite{sinha2007cache-efficient}, heavily repaired and modified to work with 64-bit pointers.              \\
sCPL-burstsort   & Copy-Burstsort with sampling, pointers and only limited copying depth by Ranjan Sinha \cite{sinha2007cache-efficient}, also heavily repaired.                 \\
\end{tabular}
\end{table}

Of the 203 different sequential string sorting variants, we selected the
twelve implementations listed in Table~\ref{tab:seqalgo} to represent both the
fastest ones in a preliminary test and each of the basic algorithms from
Section~\ref{sec:basic-sequential}. The twelve algorithms were run on all our five test
platforms on small portions of the test instances described in
Section~\ref{sec:experiments}. Tables \ref{tab:seqalgo-results1} and
\ref{tab:seqalgo-results2} show the results, with the fastest algorithm's time
highlighted with bold text.

Cells in the tables without value indicate a program error, out-of-memory
exceptions or extremely long runtime. This was always the case for the
copy-burstsort variants on the GOV2 and Wikipedia inputs, because they perform
excessive caching of characters. On Inteli7, some implementations required more
memory than the available 12\,GiB to sort the 4\,GiB prefixes of Random and URLs.

Over all run instances and platforms, multikey quicksort with caching of eight
characters was fastest on 18~pairs, winning the most tests. It was fastest on
all platforms for both URL list and GOV2 prefixes, except URL on IntelX5, and on all large instances on
AMD48 and AMD16. However, for the NoDup input, short strings with large
alphabet, the highly tuned radix sort radixR\_CE7 consistently outperformed
mkqs\_cache8 on all platforms by a small margin. The copy-burstsort variant
fbC\_burstsort was most efficient on all platforms for DNA, which are short
strings with small alphabet. For Random strings and Wikipedia suffixes,
mkqs\_cache8 or radixR\_CE7 was fastest, depending on the platforms memory
bandwidth and sequential processing speed.  Our own \emph{sequential}
implementations of S$^5$ were never the fastest, but they consistently fall in the
middle field, without any outliers.

We also measured the peak memory usage of the sequential implementations using a
heap and stack profiling
tool\footnote{\url{http://tbingmann.de/2013/malloc_count/}, by one of the
  authors.} for the selected sequential test instances. The bottom of
Table~\ref{tab:seqalgo-results1} shows the results in MiB, excluding the string
data array and the string pointer array (we only have 64-bit systems, so
pointers are eight bytes). We must note that the profiler considers
\emph{allocated virtual memory}, which may not be identical to the amount of
physical memory actually used. From the table we plainly see, that the more
\emph{caching} an implementation does, the higher its peak memory
allocation. However, the memory usage of fbC\_burstsort is extreme, even if one
considers that the implementation can deallocate and recreate the string data
from the burst trie. The lower memory usage of fbC\_burstsort for Random is due
to the high percentage of characters stored implicitly in the trie structure.
The sCPL\_burstsort and burstsortA variants bring the memory requirement down
somewhat, but they are still high. Some radixsort variants and most notable
mkqs\_cache8 are also not particularly memory conservative, again due to
caching. Our S$^5$ implementation fares well in this comparison because it does
no caching and permutes the string pointers in-place (Note that radixsort is
used for small string subsets in sequential S$^5$. This is due to the
development history: we finished sequential S$^5$ before focusing on caching
multikey quicksort). For sorting with little extra memory, plain multikey
quicksort is still a good choice.

\section{More Results on Parallel String Sorting}\label{sec:more-exp-para}

In this section we report on experiments run on all platforms shown in
Table~\ref{tab:hardware}, which contains a wide variety of multi-core machines
of different age. The results in
Figures~\ref{fig:more-IntelE5}--\ref{fig:more-IntelX5} show that each parallel
algorithm's speedup depends highly on hardware characteristics like processor
speed, RAM and cache performance\footnote{See
  \url{http://tbingmann.de/2013/pmbw/} for parallel memory bandwidth
  experiments}, and the interconnection between sockets. However, except for
Random input, our implementations of pS$^5$ is the string sorting algorithm with
highest speedup across all platforms.

We included two further parallel algorithm implementations in the tests on
Inteli7 and IntelX5: pMKQS-SIMD is a multikey quicksort implementation from
Rantala's library, which uses SIMD instructions to perform vectorized
classification against a single pivot. We improved the code to use OpenMP tasks
for recursive sorting steps. The second implementation is a parallel 2-way
LCP-mergesort also by Rantala, which we also augmented with OpenMP
tasks. However, only recursive merges are run in parallel, the largest merge is
performed sequentially. The implementation uses insertion sort for $|\Strings| <
32$, all other sorting is done via merging. N.\ Shamsundar's parallel
LCP-mergesort also uses only 2-way merges, and is omitted from the graphs
because Rantala's version is consistently faster.

\emph{IntelE5} (Figure~\ref{fig:more-IntelE5}) is the newest machine, and most
results have already been discussed in Section~\ref{sec:experiments}. The lower
three plots show that on this platform, parallel sorting becomes less efficient
for small inputs (around 300\,MiB). Compared to the following results, we also
notice that parallel multikey quicksort is relatively fast; apparently the Sandy
Bridge processor architecture optimizes sequential memory processing very well.

\emph{AMD48} (Figure~\ref{fig:more-AMD48}) is a many-core machine with high
core count, but relatively slow RAM and a slower interconnect. 
Qualitatively, the results look 
similar to the IntelE5 platform, albeit with reduced speedups.

\emph{AMD16} (Figure~\ref{fig:more-AMD16}) is an older many-core architecture
with the slowest RAM speed and interconnect in our experiment. However, on this
machine random access and processing power (in cache) seems to be most
``balanced'' for S$^5$.

\emph{Inteli7} (Figure~\ref{fig:more-Inteli7}) is a consumer-grade, single
socket machine with fast RAM and cache hierarchy. All sorting algorithms profit
from faster random access, but the gain is highest for radix sorts. Both of the
additional implementation pMKQS-SIMD and pMergesort-2way do not show any good
speedup, probably because they are already pretty slow in sequential.

\emph{IntelX5} (Figure~\ref{fig:more-IntelX5}) is the oldest architecture, and
shows the slowest absolute speedups. Nevertheless, the S$^5$ variants yield the
best gains, even for Random input on this platform.

We included the absolute running times of all our speedup experiments in
Tables~\ref{tab:absrun-IntelE5}--\ref{tab:absrun-IntelX5b} for reference and to
show that our parallel implementations scale well both for very large instances
on many-core platforms and also for small inputs on machines with fewer cores.

\begin{table}[p]\centering
\caption{Run time of sequential algorithms on IntelE5 and AMD48 in seconds, and peak memory usage of algorithms on IntelE5}\label{tab:seqalgo-results1}
\begin{tabular}{l|>{\hfill}p{33pt}>{\hfill}p{35pt}>{\hfill}p{33pt}>{\hfill}p{48pt}*{3}{>{\hfill}p{32pt}}|}
                    & \multicolumn{4}{c|}{Our Datasets} & \multicolumn{3}{c|}{Sinha's} \\   
                    & URLs     & Random   & GOV2     & \multicolumn{1}{r|}{Wikipedia} & URLs     & DNA     & NoDup    \\ \hline
$n$                 & 66\,M    & 409\,M   & 80.2\,M  & 256\,Mi   & 10\,M    & 31.5\,M & 31.6\,M  \\
$N$                 & 4\,Gi    & 4\,Gi    & 4\,Gi    & 32\,Pi    & 304\,Mi  & 302\,Mi & 382\,Mi  \\
$\frac{D}{N}$ ($D$) & 92.6\,\% & 43.0\,\% & 69.7\,\% & (13.6\,G) & 97.5\,\% & 100\,\% & 73.4\,\% \\ \hline
                    & \multicolumn{7}{c|}{IntelE5}                                               \\ \cline{2-8}
mkqs                & 36.8          & 212           & 34.7          & 128           & 5.64          & 11.0          & 10.9          \\
mkqs\_cache8        & \textbf{16.6} & 67.8          & \textbf{17.1} & 79.2          & \textbf{2.03} & 4.64          & 6.05          \\
radix8\_CI          & 48.1          & 56.8          & 40.0          & 91.7          & 6.09          & 6.75          & 6.19          \\
radix16\_CI         & 36.9          & 71.4          & 40.0          & 88.1          & 5.42          & 5.26          & 5.84          \\
radixR\_CE7         & 37.4          & \textbf{51.6} & 34.7          & \textbf{73.5} & 4.80          & 4.64          & \textbf{4.95} \\
Seq-S$^5$-Unroll    & 31.9          & 120           & 33.0          & 105           & 4.86          & 7.01          & 7.68          \\
Seq-S$^5$-Equal     & 31.4          & 187           & 34.9          & 121           & 4.98          & 7.57          & 8.20          \\
CRadix              & 55.5          & 57.7          & 41.4          & 111           & 6.77          & 10.1          & 8.55          \\
LCPMergesort        & 25.4          & 266           & 37.6          & 165           & 4.94          & 14.3          & 16.8          \\
burstsortA          & 29.6          & 127           & 30.7          & 125           & 5.64          & 8.49          & 8.53          \\
fbC\_burstsort      & 72.0          & 72.9          &               &               & 11.2          & \textbf{3.88} & 15.3          \\
sCPL\_burstsort     & 46.8          & 119           &               &               & 10.9          & 13.9          & 24.0          \\ \hline
                 & \multicolumn{7}{c|}{AMD48}                                                                                   \\ \cline{2-8}
mkqs                & 98.1          & 395           & 88.8          & 226          & 11.0          & 20.8          & 19.7          \\
mkqs\_cache8        & \textbf{34.8} & 95.9          & \textbf{33.0} & \textbf{114} & \textbf{3.47} & 7.05          & 8.76          \\
radix8\_CI          & 92.7          & 99.4          & 71.9          & 135          & 9.79          & 10.9          & 9.51          \\
radix16\_CI         & 73.8          & 135           & 61.9          & 134          & 8.77          & 9.17          & 9.45          \\
radixR\_CE7         & 85.2          & 98.6          & 66.5          & 120          & 8.27          & 8.31          & \textbf{7.66} \\
Seq-S$^5$-Unroll    & 54.9          & 203           & 54.5          & 163          & 7.66          & 11.1          & 11.7          \\
Seq-S$^5$-Equal     & 60.9          & 228           & 58.5          & 177          & 8.06          & 11.5          & 12.2          \\
CRadix              & 99.3          & \textbf{85.0} & 76.0          & 147          & 8.02          & 12.4          & 11.1          \\
LCPMergesort        & 47.1          & 452           & 73.3          & 232          & 7.33          & 20.7          & 24.5          \\
burstsortA          & 47.1          & 190           & 56.2          & 205          & 8.52          & 13.3          & 13.3          \\
fbC\_burstsort      & 98.3          & 113           &               &              & 17.4          & \textbf{5.85} & 21.8          \\
sCPL\_burstsort     & 74.4          & 203           &               &              & 19.9          & 24.7          & 37.0          \\ \hline\hline
                 & \multicolumn{7}{p{68ex}|}{\centering{}Memory usage of sequential algorithms (on IntelE5) in MiB, excluding input and string pointer array} \\ \cline{2-8}
mkqs                & 0.134   & 0.003  & 1.66   & 0.141  & 0.015  & 0.003  & 0.004  \\
mkqs\_cache8        & 1\,002  & 6\,242 & 1\,225 & 4\,096 & 153    & 483    & 483    \\
radix8\_CI          & 62.7    & 390    & 77.6   & 256    & 9.55   & 30.2   & 30.2   \\
radix16\_CI         & 126     & 781    & 155    & 513    & 20.1   & 61.3   & 61.3   \\
radixR\_CE7         & 669     & 3\,902 & 786    & 2\,567 & 111    & 303    & 303    \\
Seq-S$^5$-Unroll    & 129     & 781    & 155    & 513    & 20.3   & 60.8   & 60.9   \\
Seq-S$^5$-Equal     & 131     & 781    & 156    & 513    & 20.8   & 60.8   & 61.0   \\
CRadix              & 752     & 4\,681 & 919    & 3\,072 & 114    & 362    & 362    \\
LCPMergesort        & 1\,002  & 6\,242 & 1\,225 & 4\,096 & 153    & 483    & 483    \\
burstsortA          & 1\,466  & 7\,384 & 1\,437 & 5\,809 & 200    & 531    & 792    \\
fbC\_burstsort      & 31\,962 & 6\,200 &        &        & 2\,875 & 436    & 4\,182 \\
sCPL\_burstsort     & 9\,971  & 7\,262 &        &        & 1\,577 & 1\,697 & 6\,108 \\ \hline
\end{tabular}
\end{table}

\begin{table}[p]\centering
\caption{Run time of sequential algorithms on AMD16, Inteli7, and IntelX5 in seconds}\label{tab:seqalgo-results2}
\begin{tabular}{l|>{\hfill}p{33pt}>{\hfill}p{35pt}>{\hfill}p{33pt}>{\hfill}p{48pt}*{3}{>{\hfill}p{32pt}}|}
                    & \multicolumn{4}{c|}{Our Datasets} & \multicolumn{3}{c|}{Sinha's} \\   
                    & URLs     & Random   & GOV2     & \multicolumn{1}{r|}{Wikipedia} & URLs     & DNA     & NoDup    \\ \hline
$n$                 & 66\,M    & 409\,M   & 80.2\,M  & 256\,Mi   & 10\,M        & 31.5\,M & 31.6\,M  \\
$N$                 & 4\,Gi    & 4\,Gi    & 4\,Gi    & 32\,Pi    & 304\,Mi      & 302\,Mi & 382\,Mi  \\
$\frac{D}{N}$ ($D$) & 92.6\,\% & 43.0\,\% & 69.7\,\% & (13.6\,G) & 97.5\,\%     & 100\,\% & 73.4\,\% \\ \hline
                    & \multicolumn{7}{c|}{AMD16}                                                    \\ \cline{2-8}
mkqs                & 121           & 561          & 96.3          & 272          & 14.5          & 26.7          & 25.0          \\
mkqs\_cache8        & \textbf{43.1} & \textbf{110} & \textbf{37.5} & \textbf{135} & \textbf{4.74} & 9.03          & 10.5          \\
radix8\_CI          & 101           & 156          & 76.7          & 148          & 11.2          & 13.0          & 11.1          \\
radix16\_CI         & 81.6          & 200          & 67.1          & 146          & 10.2          & 11.4          & 10.8          \\
radixR\_CE7         & 89.8          & 155          & 73.8          & 136          & 10.1          & 10.6          & \textbf{9.32} \\
Seq-S$^5$-Unroll    & 62.3          & 288          & 57.1          & 195          & 8.50          & 12.3          & 12.5          \\
Seq-S$^5$-Equal     & 67.5          & 313          & 61.0          & 211          & 9.00          & 12.4          & 12.9          \\
CRadix              & 120           & 122          & 82.1          & 181          & 11.0          & 18.1          & 14.0          \\
LCPMergesort        & 61.7          & 681          & 92.3          & 292          & 11.6          & 32.4          & 33.1          \\
burstsortA          & 49.7          & 289          & 62.4          & 252          & 9.88          & 17.0          & 16.8          \\
fbC\_burstsort      & 124           & 171          &               &              & 25.1          & \textbf{6.22} & 29.7          \\
sCPL\_burstsort     & 83.7          & 307          &               &              & 29.0          & 41.0          & 55.2          \\ \hline
                 & \multicolumn{7}{c|}{Inteli7}                                                                               \\ \cline{2-8}
mkqs                & 33.8          & 185           & 30.8          & 111           & 5.03          & 9.33          & 9.59          \\
mkqs\_cache8        & \textbf{16.2} &               & \textbf{16.0} & 73.6          & \textbf{1.95} & 4.32          & 5.65          \\
radix8\_CI          & 40.3          & 50.1          & 33.6          & 76.0          & 5.14          & 5.58          & 5.28          \\
radix16\_CI         & 32.0          & 69.2          & 29.3          & 74.6          & 4.59          & 4.72          & 5.16          \\
radixR\_CE7         & 32.8          & \textbf{46.3} & 30.2          & \textbf{62.4} & 4.04          & 3.83          & \textbf{4.03} \\
Seq-S$^5$-Unroll    & 27.2          & 113           & 27.8          & 91.7          & 4.14          & 6.09          & 6.83          \\
Seq-S$^5$-Equal     & 26.9          & 133           & 28.7          & 101           & 4.26          & 6.38          & 7.12          \\
CRadix              & 46.6          & 145           & 35.7          & 91.4          & 5.59          & 8.19          & 6.87          \\
LCPMergesort        & 23.4          &               & 33.3          & 142           & 4.37          & 12.4          & 14.6          \\
burstsortA          & 23.1          &               & 25.2          & 106           & 4.62          & 6.79          & 7.17          \\
fbC\_burstsort      &               &               &               &               & 9.77          & \textbf{3.23} & 13.0          \\
sCPL\_burstsort     &               &               &               &               & 9.84          & 12.0          & 20.0          \\ \hline
                 & \multicolumn{7}{c|}{IntelX5}                                                                                 \\ \cline{2-8}
mkqs                & 78.0          & 323           & 55.1          & 151           & 7.53          & 13.9          & 14.0          \\
mkqs\_cache8        & 31.1          & \textbf{84.2} & \textbf{25.9} & 97.0          & \textbf{3.50} & 6.73          & 7.75          \\
radix8\_CI          & 70.7          & 103           & 49.2          & 93.6          & 5.95          & 7.38          & 7.00          \\
radix16\_CI         & 54.3          & 113           & 41.6          & 89.6          & 5.22          & 6.38          & 6.66          \\
radixR\_CE7         & 60.2          & 107           & 44.8          & \textbf{86.0} & 5.35          & 6.21          & \textbf{6.19} \\
Seq-S$^5$-Unroll    & 38.4          & 162           & 34.6          & 111           & 4.38          & 7.87          & 7.96          \\
Seq-S$^5$-Equal     & 38.5          & 191           & 35.9          & 125           & 4.64          & 8.30          & 8.75          \\
CRadix              & 80.2          & 92.3          & 57.7          & 145           & 8.86          & 13.0          & 11.1          \\
LCPMergesort        & 37.5          & 459           & 56.9          & 238           & 7.99          & 22.6          & 24.7          \\
burstsortA          & \textbf{29.0} & 215           & 34.1          & 155           & 5.88          & 9.97          & 10.8          \\
fbC\_burstsort      &               & 87.2          &               &               & 17.3          & \textbf{4.84} & 21.3          \\
sCPL\_burstsort     & 50.7          & 205           &               &               & 20.3          & 26.5          & 37.8          \\ \hline
\end{tabular}
\end{table}

\clearpage
\ifwithplots


\begin{figure}[p]\centering\parskip=\smallskipamount

\hfill%
\parbox{53mm}{\hfill\ref{speedup127}\hfill\null}
\caption{Speedup of parallel algorithm implementations on IntelE5}\label{fig:more-IntelE5}
\end{figure}


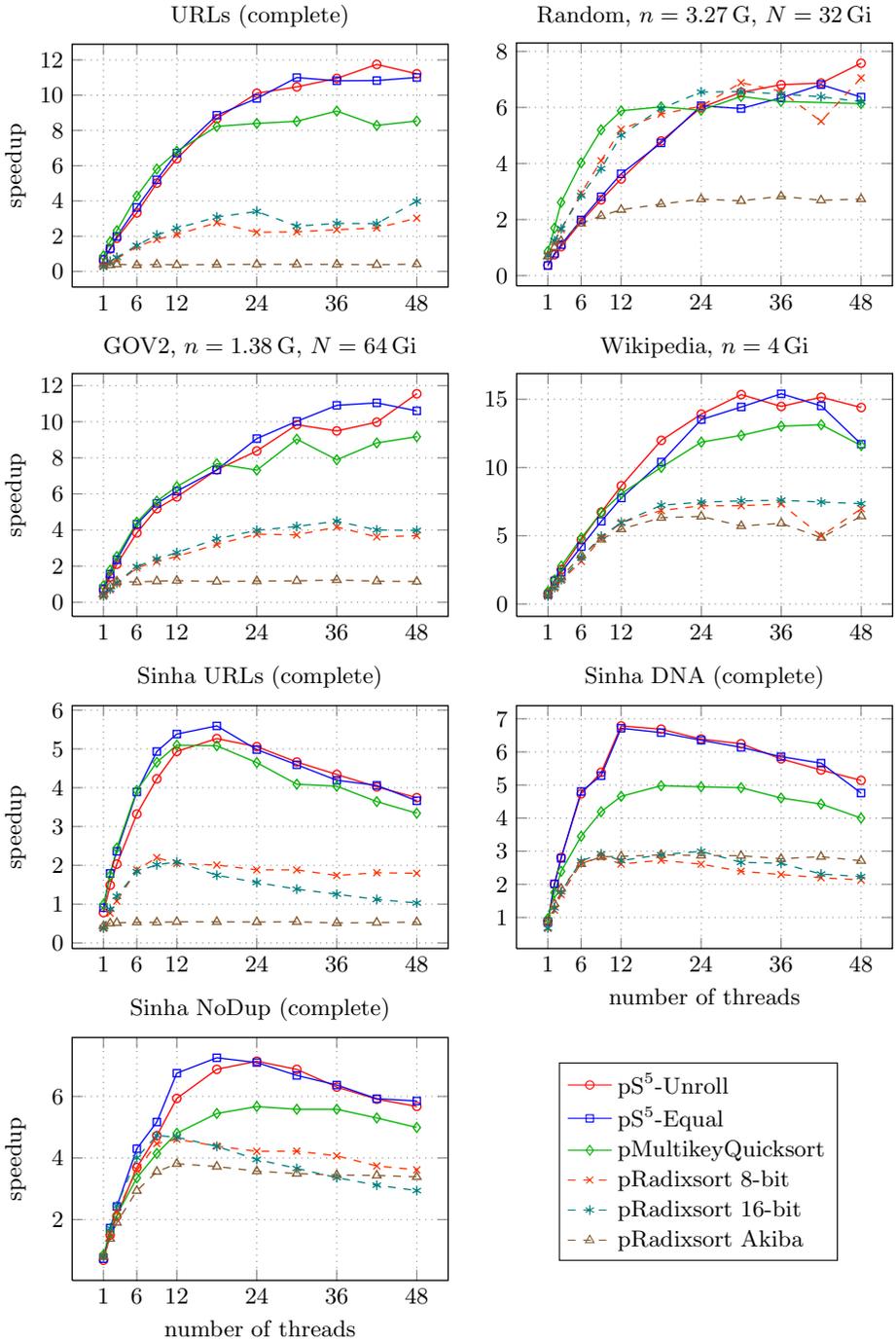
\begin{figure}[p]\centering\parskip=\smallskipamount
\begin{tikzpicture}
  \begin{axis}[plotSpeedup48,
    title={URLs (complete)},
    ylabel={speedup},
    ]

    \addplot coordinates { (1,0.64361121425919) (2,1.27079132787354) (3,1.88598118637889) (6,3.32462811819655) (9,5.01368951627483) (12,6.39390949082827) (18,8.67157970046504) (24,10.1099513239364) (30,10.4690630969369) (36,10.9530422308692) (42,11.7443641119114) (48,11.2143313518541) };
    \addlegendentry{algo=bingmann/parallel\_sample\_sortBTCU2};
    \addplot coordinates { (1,0.697466057157574) (2,1.29795598956091) (3,1.98928600121659) (6,3.63661064602086) (9,5.20204991221093) (12,6.71969682284296) (18,8.86387840866522) (24,9.81223242023027) (30,11.000871122987) (36,10.8177884712906) (42,10.8305305303742) (48,11.005341566595) };
    \addlegendentry{algo=bingmann/parallel\_sample\_sortBTCEU1};
    \addplot coordinates { (1,0.882252709565846) (2,1.68046984453246) (3,2.31339154923938) (6,4.27695078993961) (9,5.81138306789606) (12,6.81601714577844) (18,8.22630288734216) (24,8.39702681654508) (30,8.51779183012037) (36,9.09977227534339) (42,8.28468422392253) (48,8.53344460199299) };
    \addlegendentry{algo=bingmann/parallel\_mkqs};
    \addplot coordinates { (1,0.281093726560251) (2,0.511123398357441) (3,0.722106436275199) (6,1.41471775962584) (9,1.81827194167246) (12,2.09438505016207) (18,2.76830870344712) (24,2.22198078315995) (30,2.25245942117506) (36,2.37119805187032) (42,2.46698051104682) (48,3.0237126396176) };
    \addlegendentry{algo=bingmann/parallel\_radix\_sort\_8bit};
    \addplot coordinates { (1,0.263563304035765) (2,0.577584684339437) (3,0.800061392454729) (6,1.47887798634812) (9,2.06989846033504) (12,2.45117609415824) (18,3.08781571926762) (24,3.40462791085968) (30,2.57365164097215) (36,2.72823075712262) (42,2.71583359448449) (48,3.98092503804083) };
    \addlegendentry{algo=bingmann/parallel\_radix\_sort\_16bit};
    \addplot coordinates { (1,0.398028510244972) (2,0.359592533233749) (3,0.408141663772245) (6,0.362584592856022) (9,0.398896458079216) (12,0.372230544146466) (18,0.39334048190447) (24,0.403878597227077) (30,0.401984124683276) (36,0.401170010241929) (42,0.381536717442987) (48,0.416875916973327) };
    \addlegendentry{algo=akiba/parallel\_radix\_sort};

    \legend{}
  \end{axis}
\end{tikzpicture}
\hfill%
\begin{tikzpicture}
  \begin{axis}[plotSpeedup48,
    title={Random, $n = 3.27\,\text{G}$, $N = 32\,\text{Gi}$},
    ]

    \addplot coordinates { (1,0.372197964577827) (2,0.742390428985363) (3,1.03576463174916) (6,1.920954433061) (9,2.70413458572491) (12,3.4491035749001) (18,4.80072049436006) (24,5.99288912122674) (30,6.53693252430733) (36,6.81093329880339) (42,6.86742648244718) (48,7.57644337496399) };
    \addlegendentry{algo=bingmann/parallel\_sample\_sortBTCU2};
    \addplot coordinates { (1,0.358759689300445) (2,0.789806211136235) (3,1.10396046327013) (6,1.98936778476605) (9,2.81719769504384) (12,3.63381340572348) (18,4.73410383202211) (24,6.05698744826558) (30,5.95965742468675) (36,6.34305637329664) (42,6.81207481049785) (48,6.36740201663439) };
    \addlegendentry{algo=bingmann/parallel\_sample\_sortBTCEU1};
    \addplot coordinates { (1,0.863048220214655) (2,1.69575524028606) (3,2.61412325323043) (6,4.02428014653597) (9,5.19988020678867) (12,5.87886418324545) (18,6.02016400506288) (24,5.89801989340932) (30,6.39111292178599) (36,6.21243616910411) (48,6.13402160864346) };
    \addlegendentry{algo=bingmann/parallel\_mkqs};
    \addplot coordinates { (1,0.723183119550976) (2,1.21247667758662) (3,1.64708081902109) (6,2.94092090116757) (9,4.09871059712234) (12,5.21811720218018) (18,5.76032647907648) (24,6.04319235500047) (30,6.88164355307916) (36,6.57413098459005) (42,5.50143659597507) (48,7.05205918074418) };
    \addlegendentry{algo=bingmann/parallel\_radix\_sort\_8bit};
    \addplot coordinates { (1,0.716008327661449) (2,1.25219613636841) (3,1.67055634333029) (6,2.8565993131539) (9,3.81814649393238) (12,5.00687040852894) (18,5.95703702716747) (24,6.5474148977455) (30,6.56462305359988) (36,6.47567422728194) (42,6.37944023914331) (48,6.20325637539196) };
    \addlegendentry{algo=bingmann/parallel\_radix\_sort\_16bit};
    \addplot coordinates { (1,0.684272672314179) (2,1.00938395302266) (3,1.23869374230307) (6,1.86064518008486) (9,2.1329575596817) (12,2.35112522776692) (18,2.55340468426072) (24,2.73054208552688) (30,2.66927070117883) (36,2.83162331452313) (42,2.68849989627086) (48,2.73369754630111) };
    \addlegendentry{algo=akiba/parallel\_radix\_sort};

    \legend{}
  \end{axis}
\end{tikzpicture}

\begin{tikzpicture}
  \begin{axis}[plotSpeedup48,
    title={GOV2, $n = 1.38\,\text{G}$, $N = 64\,\text{Gi}$},
    ylabel={speedup},
    ]

    \addplot coordinates { (1,0.683132771962223) (2,1.40163669424006) (3,2.10625452564999) (6,3.83831930119381) (9,5.1824134491659) (12,5.83280202923886) (18,7.3327106508863) (24,8.37985642333816) (30,9.83601179767081) (36,9.49174452656533) (42,9.97578795406746) (48,11.5506459610675) };
    \addlegendentry{algo=bingmann/parallel\_sample\_sortBTCU2};
    \addplot coordinates { (1,0.755618685396913) (2,1.55895677587218) (3,2.35295358066345) (6,4.32454429358358) (9,5.47558061895147) (12,6.15324046032707) (18,7.32810809355907) (24,9.06740610583501) (30,10.0230160950684) (36,10.9111180219808) (42,11.0475889045787) (48,10.6038224379726) };
    \addlegendentry{algo=bingmann/parallel\_sample\_sortBTCEU1};
    \addplot coordinates { (1,0.913645357093959) (2,1.77588676537734) (3,2.52779015701829) (6,4.4300367730808) (9,5.60116030283081) (12,6.40355432624914) (18,7.67262340227185) (24,7.31662300611749) (30,9.03470511404664) (36,7.89027506004767) (42,8.82629528051714) (48,9.17071205586612) };
    \addlegendentry{algo=bingmann/parallel\_mkqs};
    \addplot coordinates { (1,0.334412416415689) (2,0.688330639958093) (3,1.03145651959781) (6,1.90550751253489) (9,2.258634306269) (12,2.53271887000268) (18,3.20530534207354) (24,3.77074273304932) (30,3.7310570139616) (36,4.16788419499232) (42,3.61893545865877) (48,3.6892961288714) };
    \addlegendentry{algo=bingmann/parallel\_radix\_sort\_8bit};
    \addplot coordinates { (1,0.341854590017378) (2,0.725137271028515) (3,1.06916684992617) (6,1.97497954375316) (9,2.4055167782863) (12,2.74834752623567) (18,3.51283017309896) (24,3.98469121925804) (30,4.19959154964338) (36,4.49261080492393) (42,4.00466566646074) (48,3.97998468006479) };
    \addlegendentry{algo=bingmann/parallel\_radix\_sort\_16bit};
    \addplot coordinates { (1,0.449255149927066) (2,0.896085362401838) (3,1.1497749954391) (6,1.1184924697434) (9,1.16678237586118) (12,1.19650849233387) (18,1.14323985676182) (24,1.17432523826284) (30,1.17981914084099) (36,1.23527127210693) (42,1.164787681524) (48,1.14594799356864) };
    \addlegendentry{algo=akiba/parallel\_radix\_sort};

    \legend{}
  \end{axis}
\end{tikzpicture}
\hfill%
\begin{tikzpicture}
  \begin{axis}[plotSpeedup48,
    title={Wikipedia, $n = 4\,\text{Gi}$},
    ]

    \addplot coordinates { (1,0.773930317930125) (2,1.74405012954148) (3,2.51857652484121) (6,4.70744741404265) (9,6.72949950713263) (12,8.65920784970683) (18,11.9913335625603) (24,13.9157440303064) (30,15.3457105595199) (36,14.4795882126368) (42,15.1568783512064) (48,14.4028580811448) };
    \addlegendentry{algo=bingmann/parallel\_sample\_sortBTCU2};
    \addplot coordinates { (1,0.696116838735898) (2,1.68908391734891) (3,2.33023933536738) (6,4.20644204988847) (9,6.0618118002823) (12,7.76604735394728) (18,10.399212496587) (24,13.5160627568173) (30,14.4410131558598) (36,15.4068885494233) (42,14.5201906195134) (48,11.7178423236515) };
    \addlegendentry{algo=bingmann/parallel\_sample\_sortBTCEU1};
    \addplot coordinates { (1,0.92707002187483) (2,1.76728453854983) (3,2.7823328687461) (6,4.84612474091833) (9,6.70553868321623) (12,8.08762101675025) (18,10.0125217054424) (24,11.8618507863916) (30,12.3577805006126) (36,13.0329854388603) (42,13.13702975869) (48,11.6091410053823) };
    \addlegendentry{algo=bingmann/parallel\_mkqs};
    \addplot coordinates { (1,0.574746193145378) (2,1.16421524266885) (3,1.74284252039618) (6,3.12518556666526) (9,4.79975856945771) (12,5.96759529371017) (18,6.85211355959085) (24,7.19525321096812) (30,7.20203824698816) (36,7.32791574896838) (42,4.99262644991592) (48,6.98419592230667) };
    \addlegendentry{algo=bingmann/parallel\_radix\_sort\_8bit};
    \addplot coordinates { (1,0.575295637483852) (2,1.20808170181509) (3,1.77648214066528) (6,3.39608651068712) (9,4.97014922810385) (12,5.94863532395258) (18,7.23856299045228) (24,7.46132848043676) (30,7.55724852032384) (36,7.60370176761242) (42,7.46928011477703) (48,7.3510850380303) };
    \addlegendentry{algo=bingmann/parallel\_radix\_sort\_16bit};
    \addplot coordinates { (1,0.760578960713756) (2,1.41489190100743) (3,1.96361218896698) (6,3.57701594865192) (9,4.73386135310654) (12,5.48916238408587) (18,6.32746477333835) (24,6.41923508513592) (30,5.72092646358359) (36,5.91153326947859) (42,4.86465320061106) (48,6.43498984042577) };
    \addlegendentry{algo=akiba/parallel\_radix\_sort};

    \legend{}
  \end{axis}
\end{tikzpicture}

\begin{tikzpicture}
  \begin{axis}[plotSpeedup48,
    title={Sinha URLs (complete)},
    ylabel={speedup},
    ]

    \addplot coordinates { (1,0.780623965459729) (2,1.4872803362332) (3,2.03033279313788) (6,3.32139892349302) (9,4.22847369796698) (12,4.93677990118294) (18,5.26635133069516) (24,5.05851524822334) (30,4.66182058731776) (36,4.34130966277505) (42,4.02180962139587) (48,3.74310226627428) };
    \addlegendentry{algo=bingmann/parallel\_sample\_sortBTCU2};
    \addplot coordinates { (1,0.908915034020834) (2,1.79151410103636) (3,2.35964867778553) (6,3.89225947768891) (9,4.93306888793729) (12,5.38330161306774) (18,5.59328912933353) (24,4.98367651165547) (30,4.59156116633524) (36,4.19123945764048) (42,4.05947166962414) (48,3.66350456592906) };
    \addlegendentry{algo=bingmann/parallel\_sample\_sortBTCEU1};
    \addplot coordinates { (1,0.989989503945) (2,1.73587273448252) (3,2.44580799947415) (6,3.92985483766205) (9,4.65608631127417) (12,5.09740877987301) (18,5.08163407449289) (24,4.64655369954926) (30,4.08916548573903) (36,4.03851695400443) (42,3.64150869535432) (48,3.3417724929499) };
    \addlegendentry{algo=bingmann/parallel\_mkqs};
    \addplot coordinates { (1,0.377402295492995) (2,0.765003268954798) (3,1.07639013439943) (6,1.88152395688746) (9,2.19868938865713) (12,2.04794403685378) (18,2.00707695378134) (24,1.88294739185879) (30,1.88136223098062) (36,1.73785498609363) (42,1.80825612503007) (48,1.79391090540931) };
    \addlegendentry{algo=bingmann/parallel\_radix\_sort\_8bit};
    \addplot coordinates { (1,0.376720826999273) (2,0.870065308737528) (3,1.20253632965822) (6,1.83707898452697) (9,2.01456403981559) (12,2.08637801103485) (18,1.74373093270975) (24,1.55785873082994) (30,1.3858547786073) (36,1.25593805587527) (42,1.120605095973) (48,1.02994822482136) };
    \addlegendentry{algo=bingmann/parallel\_radix\_sort\_16bit};
    \addplot coordinates { (1,0.433769111817429) (2,0.503908035392046) (3,0.517409178797366) (6,0.531484363338744) (9,0.529502420587373) (12,0.543427954278754) (18,0.543895422772998) (24,0.537952690789935) (30,0.544523818060355) (36,0.513193853970776) (42,0.524369698161148) (48,0.536534335382805) };
    \addlegendentry{algo=akiba/parallel\_radix\_sort};

    \legend{}
  \end{axis}
\end{tikzpicture}
\hfill%
\begin{tikzpicture}
  \begin{axis}[plotSpeedup48,
    title={Sinha DNA (complete)},
    xlabel={number of threads},
    every axis x label/.append style={overlay},
    ]

    \addplot coordinates { (1,0.895589765038661) (2,2.0012945227186) (3,2.81405141258086) (6,4.73395719377156) (9,5.38200912219626) (12,6.78647182845251) (18,6.68773979887185) (24,6.39011757038358) (30,6.24765657751321) (36,5.79019837908822) (42,5.45397854735355) (48,5.1401373401808) };
    \addlegendentry{algo=bingmann/parallel\_sample\_sortBTCU2};
    \addplot coordinates { (1,0.876370876784859) (2,2.01785900458052) (3,2.78468486480325) (6,4.80386032433687) (9,5.28038946130189) (12,6.71688690980548) (18,6.58464556492233) (24,6.35393732679599) (30,6.13996143768231) (36,5.85753870461861) (42,5.66212273183186) (48,4.75716775515676) };
    \addlegendentry{algo=bingmann/parallel\_sample\_sortBTCEU1};
    \addplot coordinates { (1,0.962333596788941) (2,1.74865534623075) (3,2.39287107728619) (6,3.45085998499541) (9,4.19179145618397) (12,4.65851859261297) (18,4.97913882503876) (24,4.94533412087948) (30,4.91895674216269) (36,4.60904418177432) (42,4.42520465943542) (48,4.00166476918781) };
    \addlegendentry{algo=bingmann/parallel\_mkqs};
    \addplot coordinates { (1,0.672488854394166) (2,1.22296186152771) (3,1.69012608615774) (6,2.63262526607725) (9,2.82437621052871) (12,2.61327366746919) (18,2.72788774467992) (24,2.60898244999781) (30,2.39533250291317) (36,2.29630896526221) (42,2.20014408626279) (48,2.13167222118919) };
    \addlegendentry{algo=bingmann/parallel\_radix\_sort\_8bit};
    \addplot coordinates { (1,0.675619829090827) (2,1.29511159930618) (3,1.7731766923108) (6,2.70723878414439) (9,2.92167264600365) (12,2.71653582059354) (18,2.89559418983298) (24,2.99659787346985) (30,2.66318560254754) (36,2.63643021663677) (42,2.31148128329805) (48,2.23467175284073) };
    \addlegendentry{algo=bingmann/parallel\_radix\_sort\_16bit};
    \addplot coordinates { (1,0.876827709882165) (2,1.41208908866608) (3,1.87148511668559) (6,2.61764339097368) (9,2.83183062783004) (12,2.84528737036641) (18,2.8892675805673) (24,2.87157531277156) (30,2.86341523398172) (36,2.77448240811243) (42,2.83470695535237) (48,2.71152408105192) };
    \addlegendentry{algo=akiba/parallel\_radix\_sort};

    \legend{}
  \end{axis}
\end{tikzpicture}

\begin{tikzpicture}
  \begin{axis}[plotSpeedup48,
    title={Sinha NoDup (complete)},
    legend to name={speedup126},legend columns=1,
    ylabel={speedup},xlabel={number of threads},
    ]

    \addplot coordinates { (1,0.683993226352713) (2,1.50141140038867) (3,2.10886522177776) (6,3.68966922471108) (9,4.72269010989011) (12,5.93444784033583) (18,6.87928937000312) (24,7.1429052631579) (30,6.87788015483166) (36,6.30493873527064) (42,5.9095319289368) (48,5.67559058445675) };
    \addlegendentry{algo=bingmann/parallel\_sample\_sortBTCU2};
    \addplot coordinates { (1,0.733096794443156) (2,1.72670088744749) (3,2.42510853127281) (6,4.30019371467909) (9,5.16736350694489) (12,6.75413642230902) (18,7.25447547883235) (24,7.09891949273862) (30,6.68199910443306) (36,6.37078704984781) (42,5.92630808662109) (48,5.85161974336817) };
    \addlegendentry{algo=bingmann/parallel\_sample\_sortBTCEU1};
    \addplot coordinates { (1,0.855404043325975) (2,1.57439151268262) (3,2.14632567691207) (6,3.35356213155655) (9,4.14500141457267) (12,4.80280011205216) (18,5.44531767377113) (24,5.66987989107873) (30,5.58267574701575) (36,5.5822309970385) (42,5.30022429340998) (48,4.99112100762782) };
    \addlegendentry{algo=bingmann/parallel\_mkqs};
    \addplot coordinates { (1,0.810038209491756) (2,1.56319945992818) (3,2.24678377248013) (6,3.61112094520864) (9,4.46598589500229) (12,4.61049794881478) (18,4.38558183960531) (24,4.21608088841219) (30,4.22126007763512) (36,4.06954716819143) (42,3.7400165230952) (48,3.61429554990906) };
    \addlegendentry{algo=bingmann/parallel\_radix\_sort\_8bit};
    \addplot coordinates { (1,0.81568366646624) (2,1.72054024103949) (3,2.44933075778554) (6,4.00814253667126) (9,4.7303695963557) (12,4.67497454551784) (18,4.37598394730197) (24,3.95178755333235) (30,3.66063004108964) (36,3.36051295417818) (42,3.11686970446541) (48,2.94142941360058) };
    \addlegendentry{algo=bingmann/parallel\_radix\_sort\_16bit};
    \addplot coordinates { (1,0.786358424373131) (2,1.38205221656712) (3,1.90329399871509) (6,2.9355839954535) (9,3.55604638077691) (12,3.81174684190983) (18,3.72234258288333) (24,3.57174549492147) (30,3.49437883101368) (36,3.44544663529683) (42,3.43414760959062) (48,3.38568808572947) };
    \addlegendentry{algo=akiba/parallel\_radix\_sort};

    \SpeedupLegend
  \end{axis}
\end{tikzpicture}
\hfill%
\parbox{53mm}{\hfill\ref{speedup126}\hfill\null}
\caption{Speedup of parallel algorithm implementations on AMD48}\label{fig:more-AMD48}
\end{figure}


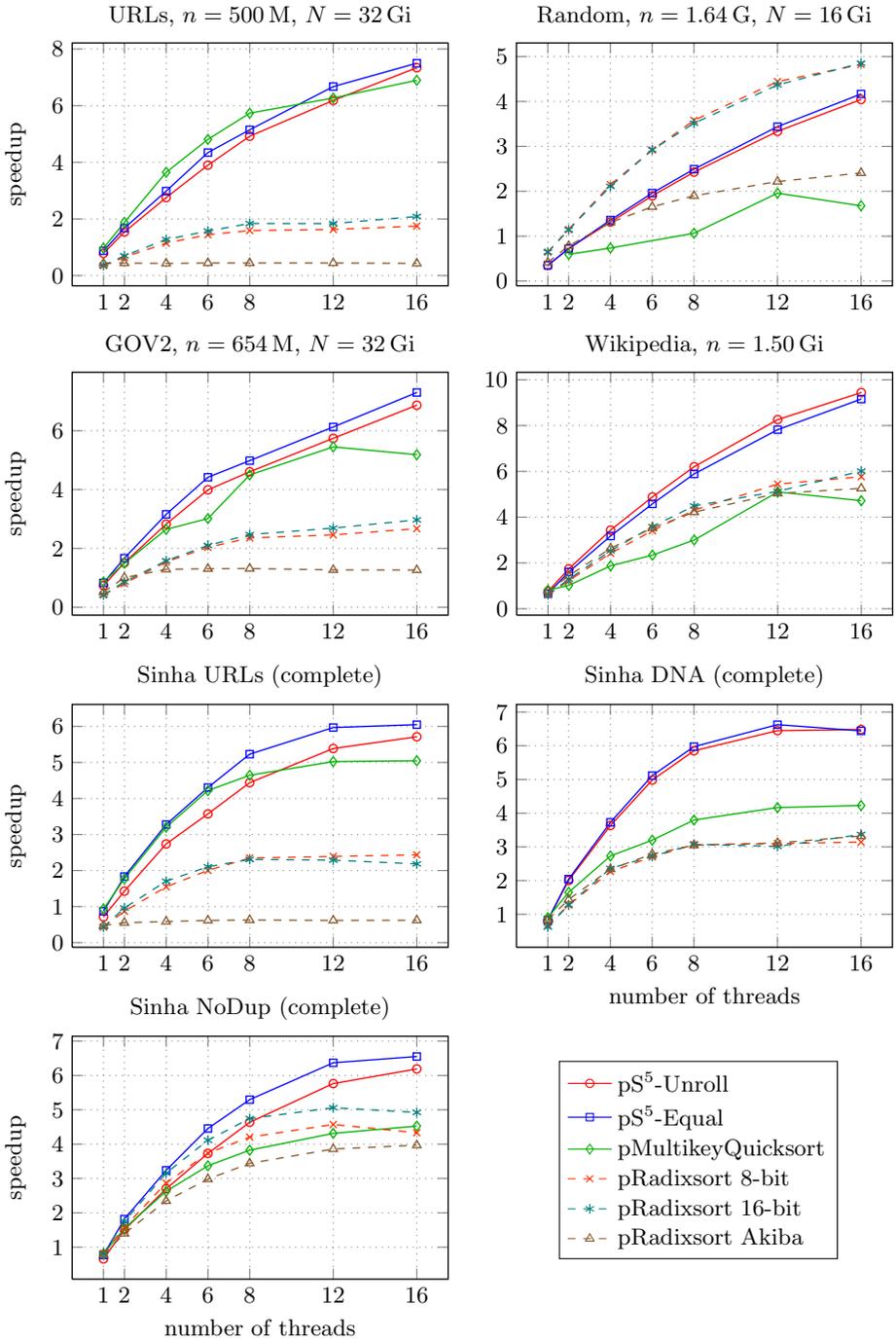
\begin{figure}[p]\centering\parskip=\smallskipamount
\begin{tikzpicture}
  \begin{axis}[plotSpeedup16,
    title={URLs, $n = 500\,\text{M}$, $N = 32\,\text{Gi}$},
    ylabel={speedup},
    ]

    \addplot coordinates { (1,0.784954672565786) (2,1.53831003057025) (4,2.75386598768814) (6,3.89889117043121) (8,4.91882948665695) (12,6.18441147476198) (16,7.33962118283726) };
    \addlegendentry{algo=bingmann/parallel\_sample\_sortBTCU2};
    \addplot coordinates { (1,0.874506681295647) (2,1.67554731070619) (4,2.98330353900729) (6,4.34280666514866) (8,5.14765029508371) (12,6.67250273712225) (16,7.49902580647426) };
    \addlegendentry{algo=bingmann/parallel\_sample\_sortBTCEU1};
    \addplot coordinates { (1,0.968035112206794) (2,1.87771292446445) (4,3.64603922371857) (6,4.80861050143508) (8,5.73290177654859) (12,6.27014912887077) (16,6.89448661838181) };
    \addlegendentry{algo=bingmann/parallel\_mkqs};
    \addplot coordinates { (1,0.353183204781627) (2,0.648261221863781) (4,1.14704366176763) (6,1.4426562209803) (8,1.59076048911991) (12,1.62964418132715) (16,1.74721081647009) };
    \addlegendentry{algo=bingmann/parallel\_radix\_sort\_8bit};
    \addplot coordinates { (1,0.353133575797653) (2,0.698331409965118) (4,1.27461691767907) (6,1.57387104617493) (8,1.83932131002148) (12,1.83713039888539) (16,2.08748741803399) };
    \addlegendentry{algo=bingmann/parallel\_radix\_sort\_16bit};
    \addplot coordinates { (1,0.432889251863906) (2,0.441667958936824) (4,0.429460637223119) (6,0.443605111176181) (8,0.44395407296409) (12,0.443727669297881) (16,0.430228295787555) };
    \addlegendentry{algo=akiba/parallel\_radix\_sort};

    \legend{}
  \end{axis}
\end{tikzpicture}
\hfill%
\begin{tikzpicture}
  \begin{axis}[plotSpeedup16,
    title={Random, $n = 1.64\,\text{G}$, $N = 16\,\text{Gi}$},
    ]

    \addplot coordinates { (1,0.344603873052314) (2,0.729814536085104) (4,1.31353409858204) (6,1.89805213305286) (8,2.42688689837571) (12,3.33426573426573) (16,4.04450875704805) };
    \addlegendentry{algo=bingmann/parallel\_sample\_sortBTCU2};
    \addplot coordinates { (1,0.353318609787284) (2,0.720864083526888) (4,1.35881825369994) (6,1.96036826236275) (8,2.49331474038225) (12,3.43933693530593) (16,4.16693941548928) };
    \addlegendentry{algo=bingmann/parallel\_sample\_sortBTCEU1};
    \addplot coordinates { (2,0.594396004898546) (4,0.736430868751476) (8,1.06489161375082) (12,1.95807934807991) (16,1.67719708452096) };
    \addlegendentry{algo=bingmann/parallel\_mkqs};
    \addplot coordinates { (1,0.64523986737939) (2,1.14667270258035) (4,2.14881057580149) (6,2.92104219971891) (8,3.57897370609403) (12,4.44300709841862) (16,4.82543210366519) };
    \addlegendentry{algo=bingmann/parallel\_radix\_sort\_8bit};
    \addplot coordinates { (1,0.645430826031909) (2,1.14261548654478) (4,2.10909406305564) (6,2.92365866121615) (8,3.51401382957969) (12,4.3690853718793) (16,4.85273251579042) };
    \addlegendentry{algo=bingmann/parallel\_radix\_sort\_16bit};
    \addplot coordinates { (1,0.413631917807054) (2,0.782144683643241) (4,1.30462634597535) (6,1.65526254884076) (8,1.89930485042568) (12,2.21726999179356) (16,2.40847434964121) };
    \addlegendentry{algo=akiba/parallel\_radix\_sort};

    \legend{}
  \end{axis}
\end{tikzpicture}

\begin{tikzpicture}
  \begin{axis}[plotSpeedup16,
    title={GOV2, $n = 654\,\text{M}$, $N = 32\,\text{Gi}$},
    ylabel={speedup},
    ]

    \addplot coordinates { (1,0.741640169782903) (2,1.51330774872239) (4,2.82636266786488) (6,3.98992467033681) (8,4.60325966355211) (12,5.74134433344228) (16,6.86679485012396) };
    \addlegendentry{algo=bingmann/parallel\_sample\_sortBTCU2};
    \addplot coordinates { (1,0.831308299653826) (2,1.67008543348393) (4,3.15771149112579) (6,4.41705798273452) (8,4.98249056815903) (12,6.12681256185883) (16,7.29792191023326) };
    \addlegendentry{algo=bingmann/parallel\_sample\_sortBTCEU1};
    \addplot coordinates { (1,0.864726672845556) (2,1.49850337940389) (4,2.64764387597847) (6,3.01572081442016) (8,4.49634460868693) (12,5.44916981864977) (16,5.18443088777015) };
    \addlegendentry{algo=bingmann/parallel\_mkqs};
    \addplot coordinates { (1,0.420843283654567) (2,0.809203578016487) (4,1.54438029161816) (6,2.03677783928889) (8,2.35907207452439) (12,2.46207362420217) (16,2.67244569314055) };
    \addlegendentry{algo=bingmann/parallel\_radix\_sort\_8bit};
    \addplot coordinates { (1,0.421842482141544) (2,0.865409793139565) (4,1.5811217332782) (6,2.10197536852885) (8,2.47232823217659) (12,2.69170244191985) (16,2.9710128366282) };
    \addlegendentry{algo=bingmann/parallel\_radix\_sort\_16bit};
    \addplot coordinates { (1,0.512228118104445) (2,1.01074003618024) (4,1.29040049764413) (6,1.3114770636957) (8,1.31804140783951) (12,1.27162635399073) (16,1.26669070192704) };
    \addlegendentry{algo=akiba/parallel\_radix\_sort};

    \legend{}
  \end{axis}
\end{tikzpicture}
\hfill%
\begin{tikzpicture}
  \begin{axis}[plotSpeedup16,
    title={Wikipedia, $n = 1.50\,\text{Gi}$},
    ]

    \addplot coordinates { (1,0.755981785222608) (2,1.74402595984157) (4,3.43589946857756) (6,4.8824026392018) (8,6.2017456894349) (12,8.26159177088084) (16,9.44309569385437) };
    \addlegendentry{algo=bingmann/parallel\_sample\_sortBTCU2};
    \addplot coordinates { (1,0.661602036755366) (2,1.60819755769498) (4,3.17914850364658) (6,4.5800106426741) (8,5.88851526752745) (12,7.82231761869771) (16,9.15915711935275) };
    \addlegendentry{algo=bingmann/parallel\_sample\_sortBTCEU1};
    \addplot coordinates { (1,0.811899138974932) (2,1.02016678155315) (4,1.87652308910303) (6,2.34260024245045) (8,3.00202844774273) (12,5.10205626022139) (16,4.7185371240163) };
    \addlegendentry{algo=bingmann/parallel\_mkqs};
    \addplot coordinates { (1,0.603570984507497) (2,1.23165813131651) (4,2.42301118893406) (6,3.4085119410405) (8,4.31149701747592) (12,5.44416630778065) (16,5.76454829163559) };
    \addlegendentry{algo=bingmann/parallel\_radix\_sort\_8bit};
    \addplot coordinates { (1,0.614332069473482) (2,1.29384534029315) (4,2.53300967226186) (6,3.58524624065704) (8,4.49182560803633) (12,5.12721173508188) (16,6.00931923069305) };
    \addlegendentry{algo=bingmann/parallel\_radix\_sort\_16bit};
    \addplot coordinates { (1,0.768130894360402) (2,1.44079829987297) (4,2.64012983424813) (6,3.52688678285686) (8,4.2204964839919) (12,5.0376295656722) (16,5.25552913867994) };
    \addlegendentry{algo=akiba/parallel\_radix\_sort};

    \legend{}
  \end{axis}
\end{tikzpicture}

\begin{tikzpicture}
  \begin{axis}[plotSpeedup16,
    title={Sinha URLs (complete)},
    ylabel={speedup},
    ]

    \addplot coordinates { (1,0.71289610594356) (2,1.42893380599716) (4,2.73680590708013) (6,3.5725064191191) (8,4.43909077928105) (12,5.38749979691638) (16,5.71346689402193) };
    \addlegendentry{algo=bingmann/parallel\_sample\_sortBTCU2};
    \addplot coordinates { (1,0.870967318304429) (2,1.83349890062092) (4,3.2769765925067) (6,4.30325550009084) (8,5.23140172525857) (12,5.96933301651078) (16,6.04948962155496) };
    \addlegendentry{algo=bingmann/parallel\_sample\_sortBTCEU1};
    \addplot coordinates { (1,0.932204612773821) (2,1.77391071762912) (4,3.20291180239409) (6,4.21928161541104) (8,4.64464530828345) (12,5.02416839851217) (16,5.04860232166546) };
    \addlegendentry{algo=bingmann/parallel\_mkqs};
    \addplot coordinates { (1,0.433657073608774) (2,0.86986228304114) (4,1.54779356280135) (6,2.00362934736198) (8,2.34758661701616) (12,2.39203343207787) (16,2.43572428922601) };
    \addlegendentry{algo=bingmann/parallel\_radix\_sort\_8bit};
    \addplot coordinates { (1,0.432358501636309) (2,0.961326351248842) (4,1.70476324303575) (6,2.10458863102668) (8,2.30837129113771) (12,2.28948531239356) (16,2.18443519274854) };
    \addlegendentry{algo=bingmann/parallel\_radix\_sort\_16bit};
    \addplot coordinates { (1,0.473263115830336) (2,0.553066886080735) (4,0.585626623277236) (6,0.615894470236949) (8,0.628762520667634) (12,0.616578904309281) (16,0.620305083815743) };
    \addlegendentry{algo=akiba/parallel\_radix\_sort};

    \legend{}
  \end{axis}
\end{tikzpicture}
\hfill%
\begin{tikzpicture}
  \begin{axis}[plotSpeedup16,
    title={Sinha DNA (complete)},
    xlabel={number of threads},
    every axis x label/.append style={overlay},
    ]

    \addplot coordinates { (1,0.822724977545813) (2,2.00584594634933) (4,3.63436959014921) (6,4.98116474490431) (8,5.85118188809722) (12,6.44937648872075) (16,6.48138839018552) };
    \addlegendentry{algo=bingmann/parallel\_sample\_sortBTCU2};
    \addplot coordinates { (1,0.831422275206821) (2,2.04465855318903) (4,3.72740781530264) (6,5.11400041663773) (8,5.97461749387504) (12,6.62595727529222) (16,6.44170456930936) };
    \addlegendentry{algo=bingmann/parallel\_sample\_sortBTCEU1};
    \addplot coordinates { (1,0.899332766722351) (2,1.65295435263699) (4,2.73001826765676) (6,3.19857684135916) (8,3.79623833500344) (12,4.1652086373251) (16,4.22694390684583) };
    \addlegendentry{algo=bingmann/parallel\_mkqs};
    \addplot coordinates { (1,0.677817046592449) (2,1.27440194749057) (4,2.26961447194058) (6,2.70295330527373) (8,3.06593574944548) (12,3.09737764708257) (16,3.14114230925401) };
    \addlegendentry{algo=bingmann/parallel\_radix\_sort\_8bit};
    \addplot coordinates { (1,0.621932171328199) (2,1.28249872876338) (4,2.36194346733991) (6,2.72272312535953) (8,3.07431105649136) (12,3.01330906757969) (16,3.36668257512174) };
    \addlegendentry{algo=bingmann/parallel\_radix\_sort\_16bit};
    \addplot coordinates { (1,0.833296220864449) (2,1.44076612653953) (4,2.33956864882861) (6,2.78585802952952) (8,3.04927079227434) (12,3.11673045703926) (16,3.3214594326433) };
    \addlegendentry{algo=akiba/parallel\_radix\_sort};

    \legend{}
  \end{axis}
\end{tikzpicture}

\begin{tikzpicture}
  \begin{axis}[plotSpeedup16,
    title={Sinha NoDup (complete)},
    legend to name={speedup124},legend columns=1,
    ylabel={speedup},xlabel={number of threads},
    ]

    \addplot coordinates { (1,0.659674039742451) (2,1.50725150027624) (4,2.69968482026237) (6,3.72940691927512) (8,4.63601588613474) (12,5.76400192785108) (16,6.19008245630597) };
    \addlegendentry{algo=bingmann/parallel\_sample\_sortBTCU2};
    \addplot coordinates { (1,0.771332262461059) (2,1.81934820921141) (4,3.23821314596524) (6,4.45514630254837) (8,5.29526265371592) (12,6.36847428563773) (16,6.55108263138856) };
    \addlegendentry{algo=bingmann/parallel\_sample\_sortBTCEU1};
    \addplot coordinates { (1,0.809456105773528) (2,1.54564009276035) (4,2.62610931729858) (6,3.36906951280443) (8,3.82376361859489) (12,4.31406668648007) (16,4.52479589578609) };
    \addlegendentry{algo=bingmann/parallel\_mkqs};
    \addplot coordinates { (1,0.820535502384086) (2,1.58158177307165) (4,2.86936760891314) (6,3.73804309792374) (8,4.20374181767537) (12,4.57247354335139) (16,4.33187460560817) };
    \addlegendentry{algo=bingmann/parallel\_radix\_sort\_8bit};
    \addplot coordinates { (1,0.814148645119304) (2,1.75129259770233) (4,3.15745165564673) (6,4.11313139178736) (8,4.74534524192764) (12,5.06071904765707) (16,4.92311954205707) };
    \addlegendentry{algo=bingmann/parallel\_radix\_sort\_16bit};
    \addplot coordinates { (1,0.782994861152288) (2,1.39905133349667) (4,2.34812019490799) (6,2.98186465528732) (8,3.43958411339205) (12,3.85816371250487) (16,3.96929435462829) };
    \addlegendentry{algo=akiba/parallel\_radix\_sort};

    \SpeedupLegend
  \end{axis}
\end{tikzpicture}
\hfill%
\parbox{53mm}{\hfill\ref{speedup124}\hfill\null}
\caption{Speedup of parallel algorithm implementations on AMD16}\label{fig:more-AMD16}
\end{figure}


\begin{figure}[p]\centering\parskip=\smallskipamount
\begin{tikzpicture}
  \begin{axis}[plotSpeedup8,
    title={URLs, $n = 66\,\text{M}$, $N = 4\,\text{Gi}$},
    ylabel={speedup},
    ]

    \addplot coordinates { (1,0.856978338215445) (2,1.65387019061669) (3,2.3354577899221) (4,2.91350610142126) (5,3.16905549358562) (6,3.34826975549951) (7,3.55172700681345) (8,3.56059841178562) };
    \addlegendentry{algo=bingmann/parallel\_sample\_sortBTCU2};
    \addplot coordinates { (1,0.902504902710747) (2,1.79508104659854) (3,2.51986324457071) (4,3.07106171353063) (5,3.32185932220743) (6,3.58080501506506) (7,3.75510119048997) (8,3.82165620100664) };
    \addlegendentry{algo=bingmann/parallel\_sample\_sortBTCEU1};
    \addplot coordinates { (1,0.948131913090489) (2,1.74861556122655) (3,2.34984218310237) (4,2.76763992343152) (5,2.85125234311767) (6,2.92405914341901) (7,2.97634216844039) (8,3.00328148069295) };
    \addlegendentry{algo=bingmann/parallel\_mkqs};
    \addplot coordinates { (1,0.399018582971554) (2,0.728484791331272) (3,0.949330383921021) (4,1.12057222590628) (5,1.13604722821368) (6,1.16631677331224) (7,1.17964492050161) (8,1.1853034443454) };
    \addlegendentry{algo=bingmann/parallel\_radix\_sort\_8bit};
    \addplot coordinates { (1,0.39904115789163) (2,0.811528879728662) (3,1.07582424505685) (4,1.27827244933338) (5,1.31640320695923) (6,1.33745836493751) (7,1.36206245119688) (8,1.36295505910622) };
    \addlegendentry{algo=bingmann/parallel\_radix\_sort\_16bit};
    \addplot coordinates { (1,0.472635047533585) (2,0.476019026324052) (3,0.476418848475059) (4,0.476791314340802) (5,0.47592545869541) (6,0.476178315530547) (7,0.475955480120234) (8,0.475838208614618) };
    \addlegendentry{algo=akiba/parallel\_radix\_sort};
    \addplot coordinates { (1,0.477989663686776) (2,0.871304730803936) (3,0.871068464758141) (4,1.11909574431391) (5,1.46141580483586) (6,1.42087745574621) (7,1.45735897873564) (8,1.50655658537491) };
    \addlegendentry{algo=rantala/mergesort\_lcp\_2way\_parallel};
    \addplot coordinates { (1,0.470223860912314) (2,0.799395856740189) (3,1.00528910357905) (4,1.13795582804011) (5,1.14323669114434) (6,1.14736045491838) (7,1.14535146431698) (8,1.13683539247496) };
    \addlegendentry{algo=rantala/multikey\_simd\_parallel4};

    \legend{}
  \end{axis}
\end{tikzpicture}
\hfill%
\begin{tikzpicture}
  \begin{axis}[plotSpeedup8,
    title={Random, $n = 205\,\text{M}$, $N = 2\,\text{Gi}$},
    ]

    \addplot coordinates { (1,0.454137448042116) (2,1.11752728207162) (3,1.6210228630929) (4,1.82138337293892) (5,2.0248531574563) (6,2.1463433475566) (7,2.37093504426362) (8,2.27276264701956) };
    \addlegendentry{algo=bingmann/parallel\_sample\_sortBTCU2};
    \addplot coordinates { (1,0.420752861747493) (2,1.02513541351727) (3,1.48764435983432) (4,1.66543592732116) (5,1.84124880347544) (6,2.06967974217978) (7,2.20137530935657) (8,2.16459645643199) };
    \addlegendentry{algo=bingmann/parallel\_sample\_sortBTCEU1};
    \addplot coordinates { (1,0.549181846757379) (2,0.986682975910983) (3,1.51450548119436) (4,1.85753350882813) (5,2.01252091262021) (6,2.11840863086744) (7,2.22159292192641) (8,2.3395703006679) };
    \addlegendentry{algo=bingmann/parallel\_mkqs};
    \addplot coordinates { (1,0.816507451903945) (2,1.55055535317051) (3,2.12498273865797) (4,2.49353394817436) (5,2.592860457219) (6,2.68035463302021) (7,2.74232009042002) (8,2.77409258669906) };
    \addlegendentry{algo=bingmann/parallel\_radix\_sort\_8bit};
    \addplot coordinates { (1,0.816091086548131) (2,1.61420807797982) (3,2.21224738049819) (4,2.40292125114111) (5,2.59738000224619) (6,2.5924891578959) (7,2.7659324417185) (8,2.66135941549285) };
    \addlegendentry{algo=bingmann/parallel\_radix\_sort\_16bit};
    \addplot coordinates { (1,0.863211628122728) (2,1.32052729205035) (3,1.5891771229291) (4,1.75089055183581) (5,1.78258564382683) (6,1.8174038256256) (7,1.84074053626309) (8,1.85090811520989) };
    \addlegendentry{algo=akiba/parallel\_radix\_sort};
    \addplot coordinates { (1,0.191599423424114) (2,0.324972444887745) (3,0.325027356674988) (4,0.372510828153572) (5,0.471213968141801) (6,0.470612037014166) (7,0.46497487413826) (8,0.471583870515282) };
    \addlegendentry{algo=rantala/mergesort\_lcp\_2way\_parallel};
    \addplot coordinates { (1,0.153947024148492) (2,0.267513768638728) (3,0.350679629350452) (4,0.408894175984662) (5,0.409439826275432) (6,0.410428959483968) (7,0.409746765569457) (8,0.410645482443986) };
    \addlegendentry{algo=rantala/multikey\_simd\_parallel4};

    \legend{}
  \end{axis}
\end{tikzpicture}

\begin{tikzpicture}
  \begin{axis}[plotSpeedup8,
    title={GOV2, $n = 80\,\text{M}$, $N = 4\,\text{Gi}$},
    ylabel={speedup},
    ]

    \addplot coordinates { (1,0.966843465252278) (2,2.09780772620447) (3,2.79137582018707) (4,3.82939298577836) (5,3.81017375489947) (6,3.97538080716954) (7,4.08386948529412) (8,3.99313993851408) };
    \addlegendentry{algo=bingmann/parallel\_sample\_sortBTCU2};
    \addplot coordinates { (1,0.922263959478665) (2,2.05000406907039) (3,2.71930503079166) (4,3.74939788406792) (5,3.72676407784973) (6,3.86217037474331) (7,4.03694290985355) (8,4.05237070412402) };
    \addlegendentry{algo=bingmann/parallel\_sample\_sortBTCEU1};
    \addplot coordinates { (1,0.938579840339857) (2,1.74097456568694) (3,2.39930881330534) (4,2.84280315316076) (5,3.07577083193124) (6,3.24952056686873) (7,3.37472704832431) (8,3.38273453410396) };
    \addlegendentry{algo=bingmann/parallel\_mkqs};
    \addplot coordinates { (1,0.475153203839082) (2,0.926760449136143) (3,1.26774427479404) (4,1.60350259010223) (5,1.66963890093614) (6,1.73429901931553) (7,1.73815765524698) (8,1.63634605951456) };
    \addlegendentry{algo=bingmann/parallel\_radix\_sort\_8bit};
    \addplot coordinates { (1,0.475218845115739) (2,0.965154586923467) (3,1.34437982215797) (4,1.70674596632859) (5,1.79876223135714) (6,1.84674265361669) (7,1.78153708904344) (8,1.96742543605563) };
    \addlegendentry{algo=bingmann/parallel\_radix\_sort\_16bit};
    \addplot coordinates { (1,0.550270392998679) (2,1.03168777895307) (3,1.15529121164847) (4,1.17089828380688) (5,1.16314599224846) (6,1.06383035325337) (7,1.10813834669807) (8,1.06164153992679) };
    \addlegendentry{algo=akiba/parallel\_radix\_sort};
    \addplot coordinates { (1,0.459844099692253) (2,0.796455814694577) (3,0.796725573796621) (4,0.905415272801372) (5,1.06608504980289) (6,1.19553794984865) (7,1.16847341738864) (8,1.24482577482052) };
    \addlegendentry{algo=rantala/mergesort\_lcp\_2way\_parallel};
    \addplot coordinates { (1,0.458791590957655) (2,0.790358004896596) (3,1.02815656606236) (4,1.18977116781399) (5,1.19607432609265) (6,1.20723175269247) (7,1.21716203259827) (8,1.21397509164181) };
    \addlegendentry{algo=rantala/multikey\_simd\_parallel4};

    \legend{}
  \end{axis}
\end{tikzpicture}
\hfill%
\begin{tikzpicture}
  \begin{axis}[plotSpeedup8,
    title={Wikipedia, $n = 512\,\text{Mi}$},
    ]

    \addplot coordinates { (1,0.798009583017823) (2,1.90485274974976) (3,2.57356571632612) (4,3.55613528159089) (5,3.59289383439001) (6,3.92053245133316) (7,4.21091934725673) (8,4.41626639246123) };
    \addlegendentry{algo=bingmann/parallel\_sample\_sortBTCU2};
    \addplot coordinates { (1,0.743766980341502) (2,1.76266555348943) (3,2.35373518940112) (4,3.30731985189405) (5,3.31267516344185) (6,3.62641004213138) (7,3.9304483000702) (8,4.28519295560005) };
    \addlegendentry{algo=bingmann/parallel\_sample\_sortBTCEU1};
    \addplot coordinates { (1,0.751731727843594) (2,1.52000950236226) (3,2.16248055813248) (4,2.66445141785131) (5,3.00155092193693) (6,3.25244173802991) (7,3.49997209167327) (8,3.59963834153762) };
    \addlegendentry{algo=bingmann/parallel\_mkqs};
    \addplot coordinates { (1,0.818063334390081) (2,1.67557969416999) (3,2.41286454745899) (4,3.13478545610886) (5,3.37576312247645) (6,3.64266294876264) (7,3.88911596059083) (8,4.12424402868991) };
    \addlegendentry{algo=bingmann/parallel\_radix\_sort\_8bit};
    \addplot coordinates { (1,0.817424540673861) (2,1.7578698603063) (3,2.51290117680533) (4,3.3105851420486) (5,3.54042380202246) (6,3.82976852134612) (7,4.09535503422331) (8,4.33208746416111) };
    \addlegendentry{algo=bingmann/parallel\_radix\_sort\_16bit};
    \addplot coordinates { (1,0.941472518463911) (2,1.75681589586549) (3,2.4627735305496) (4,3.07629058172764) (5,3.28074943951614) (6,3.4867865354727) (7,3.67585756116434) (8,3.84373678241947) };
    \addlegendentry{algo=akiba/parallel\_radix\_sort};
    \addplot coordinates { (1,0.472500254316792) (2,0.818970983610733) (3,0.814631586058524) (4,0.899002570635117) (5,1.00453528180104) (6,1.18678623206235) (7,1.22502971983742) (8,1.27327146831279) };
    \addlegendentry{algo=rantala/mergesort\_lcp\_2way\_parallel};
    \addplot coordinates { (1,0.457390448055145) (2,0.812059520571895) (3,1.07656966263199) (4,1.26802141113711) (5,1.29215380604958) (6,1.31272079812339) (7,1.32596460770206) (8,1.3372231392269) };
    \addlegendentry{algo=rantala/multikey\_simd\_parallel4};

    \legend{}
  \end{axis}
\end{tikzpicture}

\begin{tikzpicture}
  \begin{axis}[plotSpeedup8,
    title={Sinha URLs (complete)},
    ylabel={speedup},
    ]

    \addplot coordinates { (1,0.95938473615779) (2,2.08810900109031) (3,2.73007336294658) (4,3.75711170248537) (5,3.59599017220706) (6,3.86534156333859) (7,4.08185972332107) (8,4.20176698680723) };
    \addlegendentry{algo=bingmann/parallel\_sample\_sortBTCU2};
    \addplot coordinates { (1,0.930494948677805) (2,2.02111380025148) (3,2.58503725120419) (4,3.65688284947247) (5,3.68145528348902) (6,3.71354694374044) (7,3.91633252284262) (8,4.27708517663425) };
    \addlegendentry{algo=bingmann/parallel\_sample\_sortBTCEU1};
    \addplot coordinates { (1,0.953380936476864) (2,1.76157078958392) (3,2.44783069975307) (4,2.97710498970391) (5,3.10957384635569) (6,3.21399425535325) (7,3.27087506534239) (8,3.30401958042667) };
    \addlegendentry{algo=bingmann/parallel\_mkqs};
    \addplot coordinates { (1,0.376299358295253) (2,0.796137732230759) (3,1.12765135350227) (4,1.4256384083991) (5,1.47893204250652) (6,1.52620199812676) (7,1.57548826866056) (8,1.60724149268453) };
    \addlegendentry{algo=bingmann/parallel\_radix\_sort\_8bit};
    \addplot coordinates { (1,0.376409825998741) (2,0.882558438686207) (3,1.25853696212553) (4,1.60545340345087) (5,1.65000379724404) (6,1.70277966455928) (7,1.75231992830738) (8,1.79023021391721) };
    \addlegendentry{algo=bingmann/parallel\_radix\_sort\_16bit};
    \addplot coordinates { (1,0.464543396298135) (2,0.546757057913125) (3,0.580494618452673) (4,0.601142723629662) (5,0.598778172464478) (6,0.602253634783037) (7,0.603961298193864) (8,0.604987183195987) };
    \addlegendentry{algo=akiba/parallel\_radix\_sort};
    \addplot coordinates { (1,0.463519832168827) (2,0.810403529458086) (3,0.80917442582247) (4,0.918585505067707) (5,0.975460949036198) (6,1.20342310627508) (7,1.17254778502361) (8,1.18139607947388) };
    \addlegendentry{algo=rantala/mergesort\_lcp\_2way\_parallel};
    \addplot coordinates { (1,0.46952562872713) (2,0.810662233397525) (3,1.04590129736756) (4,1.19487066674) (5,1.17999873272202) (6,1.16767437902525) (7,1.16101164057606) (8,1.15058312640486) };
    \addlegendentry{algo=rantala/multikey\_simd\_parallel4};

    \legend{}
  \end{axis}
\end{tikzpicture}
\hfill%
\begin{tikzpicture}
  \begin{axis}[plotSpeedup8,
    title={Sinha DNA (complete)},
    xlabel={number of threads},
    every axis x label/.append style={overlay},
    ]

    \addplot coordinates { (1,0.869675494894935) (2,2.3511356359592) (3,3.34828215130333) (4,4.20827554512854) (5,3.80390718332829) (6,3.96257437256459) (7,4.44044397148592) (8,4.54570549821339) };
    \addlegendentry{algo=bingmann/parallel\_sample\_sortBTCU2};
    \addplot coordinates { (1,0.80665408732307) (2,2.12236352271533) (3,3.03760559254538) (4,3.84202826424152) (5,3.69489077103085) (6,3.91285082219909) (7,4.04997619728702) (8,4.52849407032519) };
    \addlegendentry{algo=bingmann/parallel\_sample\_sortBTCEU1};
    \addplot coordinates { (1,0.822294446748343) (2,1.54089867650337) (3,2.11053139671442) (4,2.54341463737997) (5,2.67755554935239) (6,2.78010006485014) (7,2.82782103676625) (8,2.84948384701077) };
    \addlegendentry{algo=bingmann/parallel\_mkqs};
    \addplot coordinates { (1,0.673308209652582) (2,1.28543262570544) (3,1.76319641187189) (4,2.22687634397681) (5,2.28197327766428) (6,2.31376864906589) (7,2.37078436461152) (8,2.39429263209131) };
    \addlegendentry{algo=bingmann/parallel\_radix\_sort\_8bit};
    \addplot coordinates { (1,0.673450617576488) (2,1.38846046674893) (3,1.94254136373762) (4,2.38270983928261) (5,2.40036660858903) (6,2.46171564224304) (7,2.50742884030362) (8,2.55173698054356) };
    \addlegendentry{algo=bingmann/parallel\_radix\_sort\_16bit};
    \addplot coordinates { (1,0.987156364571943) (2,1.70751929221814) (3,2.2547769952017) (4,2.62501710407487) (5,2.56423742803009) (6,2.59762161869396) (7,2.57182596338823) (8,2.63967252025249) };
    \addlegendentry{algo=akiba/parallel\_radix\_sort};
    \addplot coordinates { (1,0.32832003422826) (2,0.58625486943906) (3,0.584366846944541) (4,0.685756823466541) (5,0.830479088835113) (6,0.912615077963425) (7,0.909566757263877) (8,0.922015692023021) };
    \addlegendentry{algo=rantala/mergesort\_lcp\_2way\_parallel};
    \addplot coordinates { (1,0.340648832935285) (2,0.575440917206776) (3,0.728123769445723) (4,0.821594525030086) (5,0.807983321575606) (6,0.80017893769819) (7,0.792906304990034) (8,0.786637034060412) };
    \addlegendentry{algo=rantala/multikey\_simd\_parallel4};

    \legend{}
  \end{axis}
\end{tikzpicture}

\begin{tikzpicture}
  \begin{axis}[plotSpeedup8,
    title={Sinha NoDup (complete)},
    legend to name={speedup113},legend columns=1,
    ylabel={speedup},xlabel={number of threads},
    ]

    \addplot coordinates { (1,0.639551375594504) (2,1.64546158555622) (3,2.39296559450456) (4,3.08629159822352) (5,2.92060076571001) (6,3.21811600236303) (7,3.59061812937674) (8,3.75738357559485) };
    \addlegendentry{algo=bingmann/parallel\_sample\_sortBTCU2};
    \addplot coordinates { (1,0.601229020650095) (2,1.51500784876613) (3,2.20904662112511) (4,2.85608219852915) (5,2.84182258696992) (6,3.12495668712318) (7,3.27796489697675) (8,3.64454801803258) };
    \addlegendentry{algo=bingmann/parallel\_sample\_sortBTCEU1};
    \addplot coordinates { (1,0.67706534668662) (2,1.29604051932955) (3,1.80462144237204) (4,2.25047550919687) (5,2.42578474063138) (6,2.59106739833427) (7,2.70923316221286) (8,2.81624509905971) };
    \addlegendentry{algo=bingmann/parallel\_mkqs};
    \addplot coordinates { (1,0.750040659121319) (2,1.49697164187913) (3,2.13228707208852) (4,2.68108833021183) (5,2.80689559774527) (6,2.97298683222534) (7,3.115456953134) (8,3.21689158393278) };
    \addlegendentry{algo=bingmann/parallel\_radix\_sort\_8bit};
    \addplot coordinates { (1,0.74985426403569) (2,1.64309659343952) (3,2.35809369970309) (4,2.98630963550002) (5,3.09087516469791) (6,3.29041621229381) (7,3.48014183484901) (8,3.6289702410759) };
    \addlegendentry{algo=bingmann/parallel\_radix\_sort\_16bit};
    \addplot coordinates { (1,0.824336758633157) (2,1.48591105806196) (3,2.02674290850979) (4,2.46621252081333) (5,2.55073693472864) (6,2.66917688842119) (7,2.78025354438054) (8,2.87886472701735) };
    \addlegendentry{algo=akiba/parallel\_radix\_sort};
    \addplot coordinates { (1,0.290889868472924) (2,0.508504308068395) (3,0.507956951294534) (4,0.601680051830755) (5,0.625761318395303) (6,0.762738064921116) (7,0.756083497357755) (8,0.768129355606785) };
    \addlegendentry{algo=rantala/mergesort\_lcp\_2way\_parallel};
    \addplot coordinates { (1,0.277271252156387) (2,0.486514998486515) (3,0.639366989893579) (4,0.74782448462513) (5,0.747903728138879) (6,0.748614223919279) (7,0.753055195781176) (8,0.754572270837962) };
    \addlegendentry{algo=rantala/multikey\_simd\_parallel4};

    \SpeedupLegend
  \end{axis}
\end{tikzpicture}
\hfill%
\parbox{53mm}{\hfill\ref{speedup113}\hfill\null}
\caption{Speedup of parallel algorithm implementations on Inteli7}\label{fig:more-Inteli7}
\end{figure}


\begin{figure}[p]\centering\parskip=\smallskipamount

\hfill%
\parbox{53mm}{\hfill\ref{speedup150}\hfill\null}
\caption{Speedup of parallel algorithm implementations on IntelX5}\label{fig:more-IntelX5}
\end{figure}

\fi

\begin{table}[p]\centering
\caption{Absolute run time of parallel and best sequential algorithms on IntelE5 in seconds, median of 1--3 runs}\label{tab:absrun-IntelE5}
\begin{tabular}{l|*{10}{>{\hfill}p{26pt}}|}
PEs          & 1   & 2 & 4 & 8 & 12 & 16 & 24 & 32 & 48 & 64                                                                         \\ \hline
             & \multicolumn{10}{l|}{\textbf{URLs} (complete), $n = 1.11\,\text{G}$, $N = 70.7\,\text{Gi}$, $\frac{D}{N} = 93.5\,\%$} \\ \cline{2-11}
mkqs\_cache8 & 537 &  &  &  &  &  &  &  &  &  \\
pS$^5$-Unroll & 839    & 375    & 192    & 113  & 80.4   & 66.8 & 59.7 & 54.0 & 53.1   & 54.4   \\
\end{tabular}
\end{table}


\begin{table}[p]\centering
\caption{Absolute run time of parallel and best sequential algorithms on AMD48 in seconds, median of 1--3 runs}
\begin{tabular}{l|*{11}{>{\hfill}p{23.3pt}}|}
PEs             & 1   & 2 & 3 & 6 & 9 & 12 & 18 & 24 & 36 & 42 & 48 \\ \hline
& \multicolumn{11}{l|}{\textbf{URLs} (complete), $n = 1.11\,\text{G}$, $N = 70.7\,\text{Gi}$, $\frac{D}{N} = 93.5\,\%$} \\ \cline{2-12}
mkqs\_cache8 & 693 &  &  &  &  &  &  &  &  &  &  \\
pS$^5$-Unroll & 1\,077 & 546    & 368    & 209    & 138    & 108    & 80.0   & 68.6   & 63.3   & 59.0   & 61.8   \\
pS$^5$-Equal  & 994    & 534    & 349    & 191    & 133    & 103    & 78.2   & 70.7   & 64.1   & 64.0   & 63.0   \\
pMKQS         & 786    & 413    & 300    & 162    & 119    & 102    & 84.3   & 82.6   & 76.2   & 83.7   & 81.2   \\
pRS-8bit      & 2\,466 & 1\,356 & 960    & 490    & 381    & 331    & 250    & 312    & 292    & 281    & 229    \\
pRS-16bit     & 2\,630 & 1\,200 & 867    & 469    & 335    & 283    & 225    & 204    & 254    & 255    & 174    \\
pRS-Akiba     & 1\,742 & 1\,928 & 1\,699 & 1\,912 & 1\,738 & 1\,863 & 1\,763 & 1\,717 & 1\,728 & 1\,817 & 1\,663 \\ \hline
& \multicolumn{11}{l|}{\textbf{Random}, $n = 3.27\,\text{G}$, $N = 32\,\text{Gi}$, $\frac{D}{N} = 44.9\,\%$} \\ \cline{2-12}
mkqs\_cache8 & 894 &  &  &  &  &  &  &  &  &  &  \\
pS$^5$-Unroll & 2\,402 & 1\,204 & 863 & 465 & 331 & 259 & 186 & 149 & 131 & 130 & 118 \\
pS$^5$-Equal  & 2\,492 & 1\,132 & 810 & 449 & 317 & 246 & 189 & 148 & 141 & 131 & 140 \\
pMKQS         & 1\,036 & 527    & 342 & 222 & 172 & 152 & 149 & 152 & 144 &     & 146 \\
pRS-8bit      & 1\,236 & 737    & 543 & 304 & 218 & 171 & 155 & 148 & 136 & 163 & 127 \\
pRS-16bit     & 1\,249 & 714    & 535 & 313 & 234 & 179 & 150 & 137 & 138 & 140 & 144 \\
pRS-Akiba     & 1\,307 & 886    & 722 & 481 & 419 & 380 & 350 & 327 & 316 & 333 & 327 \\ \hline
& \multicolumn{11}{l|}{\textbf{GOV2}, $n = 1.82\,\text{G}$, $N = 64\,\text{Gi}$, $\frac{D}{N} = 77.0\,\%$} \\ \cline{2-12}
mkqs\_cache8 & 681 &  &  &  &  &  &  &  &  &  &  \\
pS$^5$-Unroll & 996    & 486 & 323 & 177 & 131 & 117 & 92.8 & 81.2 & 71.7 & 68.2 & 58.9 \\
pS$^5$-Equal  & 901    & 437 & 289 & 157 & 124 & 111 & 92.9 & 75.1 & 62.4 & 61.6 & 64.2 \\
pMKQS         & 745    & 383 & 269 & 154 & 122 & 106 & 88.7 & 93.0 & 86.3 & 77.1 & 74.2 \\
pRS-8bit      & 2\,035 & 989 & 660 & 357 & 301 & 269 & 212  & 181  & 163  & 188  & 184  \\
pRS-16bit     & 1\,991 & 939 & 637 & 345 & 283 & 248 & 194  & 171  & 152  & 170  & 171  \\
pRS-Akiba     & 1\,515 & 760 & 592 & 609 & 583 & 569 & 595  & 580  & 551  & 584  & 594  \\ \hline
& \multicolumn{11}{l|}{\textbf{Wikipedia}, $n = 4\,\text{G}$, $D = 249\,\text{G}$} \\ \cline{2-12}
mkqs\_cache8 & 2\,895 &  &  &  &  &  &  &  &  &  &  \\
pS$^5$-Unroll & 3\,740 & 1\,660 & 1\,149 & 615 & 430 & 334 & 241 & 208 & 200 & 191 & 201 \\
pS$^5$-Equal  & 4\,158 & 1\,714 & 1\,242 & 688 & 478 & 373 & 278 & 214 & 188 & 199 & 247 \\
pMKQS         & 3\,122 & 1\,638 & 1\,040 & 597 & 432 & 358 & 289 & 244 & 222 & 220 & 249 \\
pRS-8bit      & 5\,036 & 2\,486 & 1\,661 & 926 & 603 & 485 & 422 & 402 & 395 & 580 & 414 \\
pRS-16bit     & 5\,032 & 2\,396 & 1\,629 & 852 & 582 & 487 & 400 & 388 & 381 & 388 & 394 \\
pRS-Akiba     & 3\,806 & 2\,046 & 1\,474 & 809 & 611 & 527 & 457 & 451 & 490 & 595 & 450 \\ \hline
& \multicolumn{11}{l|}{\textbf{Sinha NoDup} (complete), $n = 31.6\,\text{M}$, $N = 382\,\text{Mi}$, $\frac{D}{N} = 73.4\,\%$} \\ \cline{2-12}
radixR\_CE6 & 8.06 &  &  &  &  &  &  &  &  &  &  \\
pS$^5$-Unroll & 11.8 & 5.37 & 3.82 & 2.18 & 1.71 & 1.36 & 1.17 & 1.13 & 1.28 & 1.36 & 1.42 \\
pS$^5$-Equal  & 11.0 & 4.67 & 3.32 & 1.87 & 1.56 & 1.19 & 1.11 & 1.14 & 1.26 & 1.36 & 1.38 \\
pMKQS         & 9.42 & 5.12 & 3.75 & 2.40 & 1.94 & 1.68 & 1.48 & 1.42 & 1.44 & 1.52 & 1.61 \\
pRS-8bit      & 9.95 & 5.15 & 3.59 & 2.23 & 1.80 & 1.75 & 1.84 & 1.91 & 1.98 & 2.15 & 2.23 \\
pRS-16bit     & 9.88 & 4.68 & 3.29 & 2.01 & 1.70 & 1.72 & 1.84 & 2.04 & 2.40 & 2.59 & 2.74 \\
pRS-Akiba     & 10.2 & 5.83 & 4.23 & 2.74 & 2.27 & 2.11 & 2.16 & 2.26 & 2.34 & 2.35 & 2.38 \\ \hline
\end{tabular}
\end{table}


\begin{table}[p]\centering
\caption{Absolute run time of parallel and best sequential algorithms on AMD16 in seconds, median of 1--3 runs}
\begin{tabular}{l|*{7}{>{\hfill}p{33pt}}|}
PEs          & 1   & 2 & 4 & 6 & 8 & 12 & 16                                                                          \\ \hline
             & \multicolumn{7}{l|}{\textbf{URLs}, $n = 500\,\text{M}$, $N = 32\,\text{Gi}$, $\frac{D}{N} = 95.4\,\%$} \\ \cline{2-8}
mkqs\_cache8 & 427 &  &  &  &  &  &  \\
pS$^5$-Unroll & 544    & 278 & 155 & 110  & 86.9 & 69.1 & 58.2 \\
pS$^5$-Equal  & 489    & 255 & 143 & 98.4 & 83.0 & 64.0 & 57.0 \\
pMKQS         & 441    & 228 & 117 & 88.8 & 74.5 & 68.1 & 62.0 \\
pRS-8bit      & 1\,210 & 659 & 372 & 296  & 269  & 262  & 245  \\
pRS-16bit     & 1\,210 & 612 & 335 & 271  & 232  & 233  & 205  \\
pRS-Akiba     & 987    & 967 & 995 & 963  & 962  & 963  & 993  \\ \hline
& \multicolumn{7}{l|}{\textbf{Random}, $n = 1.64\,\text{G}$, $N = 16\,\text{Gi}$, $\frac{D}{N} = 44.0\,\%$} \\ \cline{2-8}
mkqs\_cache8 & 486 &  &  &  &  &  &  \\
pS$^5$-Unroll & 1\,411 & 666 & 370 & 256 & 200 & 146 & 120 \\
pS$^5$-Equal  & 1\,376 & 675 & 358 & 248 & 195 & 141 & 117 \\
pMKQS         &        & 818 & 660 &     & 457 & 248 & 290 \\
pRS-8bit      & 754    & 424 & 226 & 166 & 136 & 109 & 101 \\
pRS-16bit     & 754    & 426 & 231 & 166 & 138 & 111 & 100 \\
pRS-Akiba     & 1\,176 & 622 & 373 & 294 & 256 & 219 & 202 \\ \hline
& \multicolumn{7}{l|}{\textbf{GOV2}, $n = 654\,\text{M}$, $N = 32\,\text{Gi}$, $\frac{D}{N} = 73.3\,\%$} \\ \cline{2-8}
mkqs\_cache8 & 390 &  &  &  &  &  &  \\
pS$^5$-Unroll & 526 & 258 & 138 & 97.7 & 84.7 & 67.9 & 56.8 \\
pS$^5$-Equal  & 469 & 234 & 124 & 88.3 & 78.3 & 63.7 & 53.4 \\
pMKQS         & 451 & 260 & 147 & 129  & 86.7 & 71.6 & 75.2 \\
pRS-8bit      & 927 & 482 & 253 & 191  & 165  & 158  & 146  \\
pRS-16bit     & 924 & 451 & 247 & 186  & 158  & 145  & 131  \\
pRS-Akiba     & 761 & 386 & 302 & 297  & 296  & 307  & 308  \\ \hline
& \multicolumn{7}{l|}{\textbf{Wikipedia}, $n = 1.50\,\text{Gi}$, $D = 90\,\text{G}$} \\ \cline{2-8}
mkqs\_cache8 & 1\,214 &  &  &  &  &  &  \\
pS$^5$-Unroll & 1\,605 & 696    & 353 & 249 & 196 & 147 & 129 \\
pS$^5$-Equal  & 1\,834 & 755    & 382 & 265 & 206 & 155 & 132 \\
pMKQS         & 1\,495 & 1\,190 & 647 & 518 & 404 & 238 & 257 \\
pRS-8bit      & 2\,011 & 985    & 501 & 356 & 281 & 223 & 211 \\
pRS-16bit     & 1\,975 & 938    & 479 & 338 & 270 & 237 & 202 \\
pRS-Akiba     & 1\,580 & 842    & 460 & 344 & 288 & 241 & 231 \\ \hline
& \multicolumn{7}{l|}{\textbf{Sinha NoDup} (complete), $n = 31.6\,\text{M}$, $N = 382\,\text{Mi}$, $\frac{D}{N} = 73.4\,\%$} \\ \cline{2-8}
radixR\_CE7 & 9.51 &  &  &  &  &  &  \\
pS$^5$-Unroll & 14.4 & 6.31 & 3.52 & 2.55 & 2.05 & 1.65 & 1.54 \\
pS$^5$-Equal  & 12.3 & 5.23 & 2.94 & 2.13 & 1.80 & 1.49 & 1.45 \\
pMKQS         & 11.7 & 6.15 & 3.62 & 2.82 & 2.49 & 2.20 & 2.10 \\
pRS-8bit      & 11.6 & 6.01 & 3.31 & 2.54 & 2.26 & 2.08 & 2.19 \\
pRS-16bit     & 11.7 & 5.43 & 3.01 & 2.31 & 2.00 & 1.88 & 1.93 \\
pRS-Akiba     & 12.1 & 6.80 & 4.05 & 3.19 & 2.76 & 2.46 & 2.40 \\ \hline
\end{tabular}
\end{table}


\begin{table}[p]\centering
\caption{Absolute run time of parallel and best sequential algorithms on Inteli7 in seconds, median of ten runs, larger test instances}
\begin{tabular}{l|*{8}{>{\hfill}p{30pt}}|}
PEs          & 1   & 2 & 3 & 4 & 5 & 6 & 7 & 8                                                                                      \\ \hline
             & \multicolumn{8}{l|}{\textbf{URLs}, $n = 65.7\,\text{M}$, $N = 4\,\text{Gi}$, $\frac{D}{N} = 92.7\,\%$} \\ \cline{2-9}
mkqs\_cache8 & 16.2 &  &  &  &  &  &  &  \\
pS$^5$-Unroll & 18.9 & 9.81 & 6.95 & 5.57 & 5.12 & 4.84 & 4.57 & 4.56 \\
pS$^5$-Equal  & 18.0 & 9.04 & 6.44 & 5.28 & 4.88 & 4.53 & 4.32 & 4.24 \\
pMKQS         & 17.1 & 9.28 & 6.90 & 5.86 & 5.69 & 5.55 & 5.45 & 5.40 \\
pRS-8bit      & 40.7 & 22.3 & 17.1 & 14.5 & 14.3 & 13.9 & 13.8 & 13.7 \\
pRS-16bit     & 40.7 & 20.0 & 15.1 & 12.7 & 12.3 & 12.1 & 11.9 & 11.9 \\
pRS-Akiba     & 34.3 & 34.1 & 34.1 & 34.0 & 34.1 & 34.1 & 34.1 & 34.1 \\
pMergesort    & 33.9 & 18.6 & 18.6 & 14.5 & 11.1 & 11.4 & 11.1 & 10.8 \\
pMKQS-SIMD    & 34.5 & 20.3 & 16.1 & 14.3 & 14.2 & 14.1 & 14.2 & 14.3 \\ \hline
& \multicolumn{8}{l|}{\textbf{Random}, $n = 205\,\text{M}$, $N = 2\,\text{Gi}$, $\frac{D}{N} = 42.1\,\%$} \\ \cline{2-9}
radixR\_CE7 & 20.0 &  &  &  &  &  &  &  \\
pS$^5$-Unroll & 44.1 & 17.9 & 12.3 & 11.0 & 9.88 & 9.32 & 8.44 & 8.80 \\
pS$^5$-Equal  & 47.5 & 19.5 & 13.4 & 12.0 & 10.9 & 9.67 & 9.09 & 9.24 \\
pMKQS         & 36.4 & 20.3 & 13.2 & 10.8 & 9.94 & 9.44 & 9.00 & 8.55 \\
pRS-8bit      & 24.5 & 12.9 & 9.41 & 8.02 & 7.72 & 7.46 & 7.29 & 7.21 \\
pRS-16bit     & 24.5 & 12.4 & 9.04 & 8.33 & 7.70 & 7.72 & 7.23 & 7.52 \\
pRS-Akiba     & 23.2 & 15.1 & 12.6 & 11.4 & 11.2 & 11.0 & 10.9 & 10.8 \\
pMergesort    & 104  & 61.6 & 61.5 & 53.7 & 42.5 & 42.5 & 43.0 & 42.4 \\
pMKQS-SIMD    & 130  & 74.8 & 57.0 & 48.9 & 48.9 & 48.7 & 48.8 & 48.7 \\ \hline
& \multicolumn{8}{l|}{\textbf{GOV2}, $n = 80\,\text{M}$, $N = 4\,\text{Gi}$, $\frac{D}{N} = 69.8\,\%$} \\ \cline{2-9}
mkqs\_cache8 & 16.0 &  &  &  &  &  &  &  \\
pS$^5$-Unroll & 16.5 & 7.62 & 5.73 & 4.18 & 4.20 & 4.02 & 3.92 & 4.01 \\
pS$^5$-Equal  & 17.3 & 7.80 & 5.88 & 4.27 & 4.29 & 4.14 & 3.96 & 3.95 \\
pMKQS         & 17.0 & 9.19 & 6.67 & 5.63 & 5.20 & 4.92 & 4.74 & 4.73 \\
pRS-8bit      & 33.7 & 17.3 & 12.6 & 9.98 & 9.58 & 9.22 & 9.20 & 9.78 \\
pRS-16bit     & 33.7 & 16.6 & 11.9 & 9.37 & 8.89 & 8.66 & 8.98 & 8.13 \\
pRS-Akiba     & 29.1 & 15.5 & 13.8 & 13.7 & 13.8 & 15.0 & 14.4 & 15.1 \\
pMergesort    & 34.8 & 20.1 & 20.1 & 17.7 & 15.0 & 13.4 & 13.7 & 12.8 \\
pMKQS-SIMD    & 34.9 & 20.2 & 15.6 & 13.4 & 13.4 & 13.2 & 13.1 & 13.2 \\ \hline
& \multicolumn{8}{l|}{\textbf{Wikipedia}, $n = 2\,\text{G}$, $D = 13.8\,\text{G}$} \\ \cline{2-9}
radixR\_CE7 & 62.7 &  &  &  &  &  &  &  \\
pS$^5$-Unroll & 78.6 & 32.9 & 24.4 & 17.6 & 17.5 & 16.0 & 14.9 & 14.2 \\
pS$^5$-Equal  & 84.3 & 35.6 & 26.6 & 19.0 & 18.9 & 17.3 & 16.0 & 14.6 \\
pMKQS         & 83.4 & 41.3 & 29.0 & 23.5 & 20.9 & 19.3 & 17.9 & 17.4 \\
pRS-8bit      & 76.7 & 37.4 & 26.0 & 20.0 & 18.6 & 17.2 & 16.1 & 15.2 \\
pRS-16bit     & 76.7 & 35.7 & 25.0 & 18.9 & 17.7 & 16.4 & 15.3 & 14.5 \\
pRS-Akiba     & 66.6 & 35.7 & 25.5 & 20.4 & 19.1 & 18.0 & 17.1 & 16.3 \\
pMergesort    & 133  & 76.6 & 77.0 & 69.7 & 62.4 & 52.8 & 51.2 & 49.2 \\
pMKQS-SIMD    & 137  & 77.2 & 58.2 & 49.5 & 48.5 & 47.8 & 47.3 & 46.9 \\ \hline
\end{tabular}
\end{table}

\begin{table}[p]\centering
\caption{Absolute run time of parallel and best sequential algorithms on Inteli7 in seconds, median of ten runs, smaller test instances}
\begin{tabular}{l|*{8}{>{\hfill}p{30pt}}|}
PEs          & 1   & 2 & 3 & 4 & 5 & 6 & 7 & 8                                                                                      \\ \hline
& \multicolumn{8}{l|}{\textbf{Sinha URLs} (complete), $n = 10\,\text{M}$, $N = 304\,\text{Mi}$, $\frac{D}{N} = 97\,5\,\%$} \\ \cline{2-9}
mkqs\_cache8 & 1.96 &  &  &  &  &  &  &  \\
pS$^5$-Unroll & 2.04 & 0.936 & 0.716 & 0.520 & 0.544 & 0.506 & 0.479 & 0.465 \\
pS$^5$-Equal  & 2.10 & 0.967 & 0.756 & 0.535 & 0.531 & 0.527 & 0.499 & 0.457 \\
pMKQS         & 2.05 & 1.11  & 0.799 & 0.657 & 0.629 & 0.608 & 0.598 & 0.592 \\
pRS-8bit      & 5.20 & 2.46  & 1.73  & 1.37  & 1.32  & 1.28  & 1.24  & 1.22  \\
pRS-16bit     & 5.19 & 2.22  & 1.55  & 1.22  & 1.19  & 1.15  & 1.12  & 1.09  \\
pRS-Akiba     & 4.21 & 3.58  & 3.37  & 3.25  & 3.27  & 3.25  & 3.24  & 3.23  \\
pMergesort    & 4.22 & 2.41  & 2.42  & 2.13  & 2.00  & 1.62  & 1.67  & 1.66  \\
pMKQS-SIMD    & 4.16 & 2.41  & 1.87  & 1.64  & 1.66  & 1.67  & 1.68  & 1.70  \\ \hline
& \multicolumn{8}{l|}{\textbf{Sinha DNA} (complete), $n = 31.6\,\text{M}$, $N = 302\,\text{Mi}$, $\frac{D}{N} = 100\,\%$} \\ \cline{2-9}
radixR\_CE6 & 3.84 &  &  &  &  &  &  &  \\
pS$^5$-Unroll & 4.41 & 1.63 & 1.15 & 0.912 & 1.01 & 0.968 & 0.864 & 0.844 \\
pS$^5$-Equal  & 4.76 & 1.81 & 1.26 & 0.999 & 1.04 & 0.981 & 0.947 & 0.847 \\
pMKQS         & 4.67 & 2.49 & 1.82 & 1.51  & 1.43 & 1.38  & 1.36  & 1.35  \\
pRS-8bit      & 5.70 & 2.98 & 2.18 & 1.72  & 1.68 & 1.66  & 1.62  & 1.60  \\
pRS-16bit     & 5.70 & 2.76 & 1.98 & 1.61  & 1.60 & 1.56  & 1.53  & 1.50  \\
pRS-Akiba     & 3.89 & 2.25 & 1.70 & 1.46  & 1.50 & 1.48  & 1.49  & 1.45  \\
pMergesort    & 11.7 & 6.54 & 6.57 & 5.60  & 4.62 & 4.20  & 4.22  & 4.16  \\
pMKQS-SIMD    & 11.3 & 6.67 & 5.27 & 4.67  & 4.75 & 4.79  & 4.84  & 4.88  \\ \hline
& \multicolumn{8}{l|}{\textbf{Sinha NoDup} (complete), $n = 31.6\,\text{M}$, $N = 382\,\text{Mi}$, $\frac{D}{N} = 73.4\,\%$} \\ \cline{2-9}
radixR\_CE6 & 4.06 &  &  &  &  &  &  &  \\
pS$^5$-Unroll & 6.35 & 2.47 & 1.70 & 1.31 & 1.39 & 1.26 & 1.13 & 1.08 \\
pS$^5$-Equal  & 6.75 & 2.68 & 1.84 & 1.42 & 1.43 & 1.30 & 1.24 & 1.11 \\
pMKQS         & 5.99 & 3.13 & 2.25 & 1.80 & 1.67 & 1.57 & 1.50 & 1.44 \\
pRS-8bit      & 5.41 & 2.71 & 1.90 & 1.51 & 1.45 & 1.37 & 1.30 & 1.26 \\
pRS-16bit     & 5.41 & 2.47 & 1.72 & 1.36 & 1.31 & 1.23 & 1.17 & 1.12 \\
pRS-Akiba     & 4.92 & 2.73 & 2.00 & 1.65 & 1.59 & 1.52 & 1.46 & 1.41 \\
pMergesort    & 14.0 & 7.98 & 7.99 & 6.75 & 6.49 & 5.32 & 5.37 & 5.28 \\
pMKQS-SIMD    & 14.6 & 8.34 & 6.35 & 5.43 & 5.43 & 5.42 & 5.39 & 5.38 \\ \hline
\end{tabular}
\end{table}


\begin{table}[p]\centering
\caption{Absolute run time of parallel and best sequential algorithms on IntelX5 in seconds, median of ten runs, larger test instances}
\begin{tabular}{l|*{8}{>{\hfill}p{30pt}}|}
PEs          & 1   & 2 & 3 & 4 & 5 & 6 & 7 & 8                                                                                      \\ \hline
             & \multicolumn{8}{l|}{\textbf{URLs}, $n = 65.7\,\text{M}$, $N = 4\,\text{Gi}$, $\frac{D}{N} = 92.7\,\%$} \\ \cline{2-9}
mkqs\_cache8 & 64.1 &  &  &  &  &  &  &  \\
pS$^5$-Unroll & 62.3 & 32.8 & 25.6 & 22.0 & 20.6 & 19.2 & 18.8 & 19.0 \\
pS$^5$-Equal  & 60.4 & 32.2 & 25.4 & 21.8 & 20.5 & 19.2 & 18.8 & 19.0 \\
pMKQS         & 67.3 & 36.7 & 30.7 & 27.2 & 26.9 & 26.9 & 26.9 & 26.7 \\
pRS-8bit      & 146  & 89.1 & 80.5 & 74.1 & 73.6 & 72.8 & 74.4 & 75.0 \\
pRS-16bit     & 146  & 78.0 & 69.0 & 63.9 & 62.6 & 62.4 & 63.9 & 62.1 \\
pRS-Akiba     & 120  & 118  & 118  & 118  & 118  & 118  & 118  & 117  \\
pMergesort    & 91.3 & 53.2 & 53.1 & 44.6 & 37.7 & 38.4 & 40.4 & 42.8 \\
pMKQS-SIMD    & 155  & 94.3 & 85.9 & 80.3 & 79.9 & 79.7 & 79.3 & 80.2 \\ \hline
& \multicolumn{8}{l|}{\textbf{Random}, $n = 307\,\text{M}$, $N = 3\,\text{Gi}$, $\frac{D}{N} = 42.8\,\%$} \\ \cline{2-9}
mkqs\_cache8 & 61.2 &  &  &  &  &  &  &  \\
pS$^5$-Unroll & 98.0 & 48.9 & 39.2 & 30.2 & 29.3 & 27.0 & 25.3 & 23.6 \\
pS$^5$-Equal  & 106  & 52.9 & 42.4 & 31.9 & 30.7 & 28.0 & 26.4 & 24.0 \\
pMKQS         & 68.2 & 39.7 & 31.2 & 27.1 & 27.3 & 26.7 & 27.2 & 26.6 \\
pRS-8bit      & 76.8 & 45.4 & 37.3 & 32.0 & 30.1 & 28.5 & 27.7 & 27.0 \\
pRS-16bit     & 76.9 & 42.5 & 34.9 & 28.7 & 27.6 & 26.0 & 25.0 & 24.3 \\
pRS-Akiba     & 78.7 & 48.5 & 40.3 & 35.7 & 33.5 & 31.9 & 31.1 & 30.5 \\
pMergesort    & 301  & 188  & 182  & 177  & 157  & 158  & 169  & 172  \\
pMKQS-SIMD    & 469  & 282  & 253  & 237  & 232  & 231  & 228  & 226  \\ \hline
& \multicolumn{8}{l|}{\textbf{GOV2}, $n = 166\,\text{M}$, $N = 8\,\text{Gi}$, $\frac{D}{N} = 70.6\,\%$} \\ \cline{2-9}
mkqs\_cache8 & 55.7 &  &  &  &  &  &  &  \\
pS$^5$-Unroll & 52.4 & 28.2 & 21.3 & 17.8 & 15.8 & 15.4 & 14.9 & 15.5 \\
pS$^5$-Equal  & 54.2 & 28.8 & 21.8 & 18.1 & 15.8 & 15.4 & 15.2 & 15.7 \\
pMKQS         & 59.4 & 33.1 & 27.2 & 24.3 & 23.6 & 23.3 & 23.7 & 23.6 \\
pRS-8bit      & 111  & 60.6 & 49.5 & 44.2 & 44.4 & 45.5 & 44.6 & 44.5 \\
pRS-16bit     & 111  & 58.4 & 46.0 & 40.4 & 38.4 & 40.6 & 39.6 & 39.4 \\
pRS-Akiba     & 93.2 & 51.4 & 47.3 & 48.3 & 48.5 & 49.8 & 51.5 & 52.8 \\
pMergesort    & 125  & 77.2 & 77.2 & 73.0 & 64.6 & 66.3 & 69.4 & 71.8 \\
pMKQS-SIMD    & 168  & 100  & 90.2 & 84.6 & 83.2 & 82.1 & 81.9 & 82.0 \\ \hline
& \multicolumn{8}{l|}{\textbf{Wikipedia}, $n = 2\,\text{G}$, $D = 13.8\,\text{G}$} \\ \cline{2-9}
 & 83.7 &  &  &  &  &  &  &  \\
pS$^5$-Unroll & 99.9 & 48.5 & 37.7 & 29.1 & 27.7 & 25.8 & 24.4 & 22.9 \\
pS$^5$-Equal  & 109  & 51.2 & 40.1 & 30.3 & 29.0 & 26.7 & 25.3 & 23.4 \\
pMKQS         & 102  & 55.2 & 44.4 & 37.7 & 36.1 & 35.3 & 34.6 & 34.5 \\
pRS-8bit      & 91.6 & 50.1 & 38.7 & 31.9 & 30.1 & 28.8 & 27.7 & 27.2 \\
pRS-16bit     & 91.8 & 48.5 & 37.7 & 30.6 & 28.9 & 27.7 & 26.3 & 25.2 \\
pRS-Akiba     & 84.7 & 47.4 & 36.8 & 31.1 & 29.1 & 28.3 & 27.3 & 27.0 \\
pMergesort    & 193  & 125  & 128  & 120  & 116  & 120  & 138  & 161  \\
pMKQS-SIMD    & 263  & 156  & 138  & 127  & 126  & 126  & 126  & 126  \\ \hline
\end{tabular}
\end{table}

\begin{table}[p]\centering
\caption{Absolute run time of parallel and best sequential algorithms on IntelX5 in seconds, median of ten runs, smaller test instances}\label{tab:absrun-IntelX5b}
\begin{tabular}{l|*{8}{>{\hfill}p{30pt}}|}
PEs          & 1   & 2 & 3 & 4 & 5 & 6 & 7 & 8                                                                                      \\ \hline
& \multicolumn{8}{l|}{\textbf{Sinha URLs} (complete), $n = 10\,\text{M}$, $N = 304\,\text{Mi}$, $\frac{D}{N} = 97\,5\,\%$} \\ \cline{2-9}
mkqs\_cache8 & 3.34 &  &  &  &  &  &  &  \\
pS$^5$-Unroll & 3.22 & 1.69 & 1.38 & 1.09 & 1.03 & 1.02 & 0.989 & 0.947 \\
pS$^5$-Equal  & 3.28 & 1.72 & 1.41 & 1.10 & 1.02 & 1.04 & 0.999 & 0.940 \\
pMKQS         & 3.59 & 2.00 & 1.64 & 1.44 & 1.42 & 1.42 & 1.43  & 1.46  \\
pRS-8bit      & 5.75 & 3.21 & 2.67 & 2.36 & 2.30 & 2.25 & 2.23  & 2.23  \\
pRS-16bit     & 5.74 & 2.86 & 2.36 & 2.05 & 2.00 & 1.97 & 1.95  & 1.94  \\
pRS-Akiba     & 5.09 & 4.34 & 4.18 & 4.08 & 4.07 & 4.07 & 4.07  & 4.07  \\
pMergesort    & 6.48 & 4.10 & 3.76 & 3.92 & 3.55 & 3.69 & 3.86  & 4.32  \\
pMKQS-SIMD    & 9.52 & 5.88 & 5.47 & 5.23 & 5.22 & 5.24 & 5.27  & 5.29  \\ \hline
& \multicolumn{8}{l|}{\textbf{Sinha DNA} (complete), $n = 31.6\,\text{M}$, $N = 302\,\text{Mi}$, $\frac{D}{N} = 100\,\%$} \\ \cline{2-9}
radixR\_CE7 & 6.08 &  &  &  &  &  &  &  \\
pS$^5$-Unroll & 5.71 & 2.85 & 2.22 & 1.85 & 1.74 & 1.59 & 1.66 & 1.47 \\
pS$^5$-Equal  & 6.07 & 3.04 & 2.34 & 1.94 & 1.80 & 1.69 & 1.74 & 1.50 \\
pMKQS         & 7.36 & 4.19 & 3.66 & 3.28 & 3.28 & 3.26 & 3.28 & 3.32 \\
pRS-8bit      & 7.46 & 4.42 & 3.97 & 3.53 & 3.53 & 3.52 & 3.52 & 3.54 \\
pRS-16bit     & 7.47 & 4.27 & 3.69 & 3.32 & 3.29 & 3.28 & 3.30 & 3.28 \\
pRS-Akiba     & 5.98 & 3.77 & 3.51 & 3.20 & 3.21 & 3.20 & 3.19 & 3.18 \\
pMergesort    & 18.4 & 10.9 & 10.7 & 9.56 & 8.30 & 8.42 & 8.68 & 9.04 \\
pMKQS-SIMD    & 27.0 & 16.8 & 15.7 & 14.9 & 14.9 & 14.9 & 15.0 & 15.0 \\ \hline
& \multicolumn{8}{l|}{\textbf{Sinha NoDup} (complete), $n = 31.6\,\text{M}$, $N = 382\,\text{Mi}$, $\frac{D}{N} = 73.4\,\%$} \\ \cline{2-9}
radixR\_CE7 & 6.06 &  &  &  &  &  &  &  \\
pS$^5$-Unroll & 7.58 & 3.80 & 2.80 & 2.26 & 2.03 & 1.84 & 1.86 & 1.67 \\
pS$^5$-Equal  & 7.94 & 3.98 & 2.91 & 2.34 & 2.08 & 1.90 & 1.93 & 1.69 \\
pMKQS         & 8.37 & 4.50 & 3.81 & 3.25 & 3.15 & 3.14 & 3.10 & 3.09 \\
pRS-8bit      & 7.19 & 4.16 & 3.37 & 2.92 & 2.79 & 2.68 & 2.66 & 2.61 \\
pRS-16bit     & 7.19 & 3.68 & 2.93 & 2.49 & 2.36 & 2.23 & 2.23 & 2.15 \\
pRS-Akiba     & 7.03 & 4.07 & 3.27 & 2.82 & 2.68 & 2.60 & 2.54 & 2.52 \\
pMergesort    & 20.8 & 12.9 & 11.8 & 11.8 & 10.4 & 10.6 & 11.0 & 11.5 \\
pMKQS-SIMD    & 30.1 & 18.0 & 16.1 & 15.0 & 14.9 & 14.9 & 14.9 & 15.1 \\ \hline
\end{tabular}
\end{table}

\end{appendix}

\end{document}